\documentclass[10pt,final,doublecolumn]{IEEEtran}
\IEEEoverridecommandlockouts
\usepackage{amsmath}
\usepackage{amssymb}
\usepackage{amsfonts}
\usepackage{latexsym}
\usepackage{graphicx}
\usepackage{bbding}
\usepackage{enumerate}
\usepackage{subeqnarray}
\usepackage{multicol}
\usepackage{color}
\usepackage{slashbox}
\usepackage{setspace}
\usepackage{mathrsfs}
\usepackage{array}
\usepackage{algorithm,algorithmic}
\usepackage{colortbl}
\usepackage{bm}
\allowdisplaybreaks[4]

\newtheorem{lemma}{Lemma}
\newtheorem{remark}{Remark}

\begin{document}
\title{Deep Learning-based Design of Uplink Integrated Sensing and Communication}
\author{Qiao Qi, Xiaoming Chen, Caijun Zhong, Chau Yuen, and Zhaoyang Zhang
\thanks{Qiao Qi is with the School of Information Science and Technology, Hangzhou Normal University, Hangzhou, 311121, China (e-mail: qiqiao@hznu.edu.cn). Xiaoming Chen, Caijun Zhong, and Zhaoyang Zhang are with the College of Information Science and Electronic Engineering, Zhejiang University, Hangzhou, 310027, China (e-mail: \{chen\_xiaoming, caijunzhong, and ning\_ming\}@zju.edu.cn). Chau Yuen is with the School of Electrical and Electronic Engineering, Nanyang Technological University, 639798, Singapore (e-mail: chau.yuen@ntu.edu.sg).}}\maketitle

\begin{abstract}
 In this paper, we investigate the issue of uplink integrated sensing and communication (ISAC) in 6G wireless networks where the sensing echo signal and the communication signal are received simultaneously at the base station (BS).  To effectively mitigate the mutual interference between sensing and communication caused by the sharing of spectrum and hardware resources, we provide a joint sensing transmit waveform and communication receive beamforming design with the objective of maximizing the weighted sum of normalized sensing rate and normalized communication rate. It is formulated as a computationally complicated non-convex optimization problem, which is quite difficult to be solved by conventional optimization methods.  To this end, we first make a series of equivalent transformation on the optimization problem to reduce the design complexity, and then develop a deep learning (DL)-based scheme to enhance the overall performance of ISAC.  Both theoretical analysis and simulation results confirm the effectiveness and robustness of the proposed DL-based scheme for ISAC in 6G wireless networks.

\end{abstract}
\begin{IEEEkeywords}
6G, integrated sensing and communication, deep learning, waveform and beamforming design.
\end{IEEEkeywords}

\IEEEpeerreviewmaketitle

\section{Introduction}

Integrated sensing and communication (ISAC) has been widely recognized as one of main use cases of the sixth-generation (6G) wireless networks, which provides wide application prospects in the fields of smart city, smart medical, automatic driving, etc. In general, ISAC integrates sensing and communication functions into an identical system with limited radio spectrum by sharing wireless and hardware resources \cite{C0}-\cite{C3}. However, it is quite challenging to achieve an appropriate performance balance between sensing and communication for ISAC in 6G wireless networks. In particular, the ISAC system may suffer from severe inter-functionality interference between sensing and communication caused by resources sharing, resulting in performance degradation. Thus, how to effectively coordinate the inter-functionality interference between sensing and communication has become one of the most important issues for ISAC systems. Traditionally, it is usual to allocate different orthogonal radio resources to sensing and communication functions, respectively. For example, the authors studied time/frequency/code-division enabled ISAC systems to avoid the inter-functionality interference \cite{C4}-\cite{C6}. Although the transmission process is interference-free, orthogonal-enabled ISAC leads to a low spectrum efficiency. Therefore, it is desired to develop non-orthogonal ISAC systems for improving the spectrum utilization \cite{C7}.

In non-orthogonal ISAC systems, the base station (BS) is a multi-functional network node with the ability to handle sensing and communication simultaneously. Most of the existing non-orthogonal ISAC works pay more attention to downlink ISAC systems, where the BS first broadcasts the sensing waveform to the targets and the communication signals to the users simultaneously, and then received the echo signals form targets for sensing information extraction \cite{C8}-\cite{C10}. In this context, there is no communication-to-sensing interference since the BS has a prior knowledge about communication message. Thus, the key of design for the downlink non-orthogonal ISAC systems is to mitigate the sensing-to-communication interference and the inter-user communication interference. To this end, joint design of signal waveform and transmit beamforming for the BS has been extensively studied.  For example, the optimal waveform design was proposed in \cite{C8} for minimizing the downlink multi-user interference for ISAC systems. In \cite{C9}, a transmit beamforming design was put forward to achieve the performance trade-off between sensing and communication for ISAC systems. The authors in \cite{C10} studied a joint beamforming design for ISAC system by maximizing communication rate while meeting the accuracy of radar beam pattern.

Compared to downlink ISAC systems, there are few literatures on uplink non-orthogonal ISAC systems, where the sensing echo signal and communication signal are received at the BS simultaneously, and the BS should recover the communication information from communication signal and estimate the targets parameters from sensing echo signal. Different from downlink non-orthogonal ISAC systems, the BS in uplink non-orthogonal ISAC systems has no prior knowledge about communication signal. That is to say, the BS needs to process the mixed received signal that exists the severe mutual interference between sensing and communication. In this context, efficient interference coordination schemes play a crucial role in unleashing the full potential of non-orthogonal uplink ISAC systems. Inspired by non-orthogonal multiple access (NOMA)-based multi-user communication systems \cite{C11}, \cite{C12}, previous works employed the successive interference cancellation (SIC) technique to decode the communication signals first and then performed sensing information estimation without communication interference \cite{C13}, \cite{C14}.  However, there exists some limitations.  The communication signal decoding is always under the sensing interference, leading to a limited communication rate.  Moreover, imperfect SIC implementation can result in residual communication interference during the sensing information extraction, leading to a low sensing rate. Although it would make sense to consider another SIC order where the sensing information is estimated first by treating the communication signal as interference, quantifying the influence of the communication signal on estimating sensing information and evaluating the statistics of the residual error becomes challenging, which makes subsequent analyses intractable. This is because the power of the sensing echo signal is typically much weaker than the communication signal. The sensing echo signal is reflected off from distant targets, experiencing double pathloss, while the communication signal is directly transmitted from the communication users (CUs). As a result, if the BS tries to decode the sensing signal first, it would be heavily distorted due to the interference from the stronger communication signal.

Considering the potential limitations and uncertainties introduced by the use of SIC technique, this paper proposes a novel framework for simultaneously implementing sensing information extraction and communication signal decoding in practical uplink ISAC systems facing mutual interference caused by resource sharing. To effectively mitigate the mutual interference between sensing and communication, it is necessary to select appropriate performance indicators for ISAC system. Note that although the sensing performance is usually evaluated by estimation theory-based metrics, such as the Cramer-Rao bound, mean squared error (MSE), detection probability, etc., they all depend on special estimation methods \cite{C15}, \cite{C16}. For example, the well-known Cramer-Rao bound was derived according to the unbiased estimation method \cite{C17}. Instead, mutual information (MI)-based metrics is more general for performance evaluation. Besides, it was validated in \cite{C18} that maximizing the sensing MI can achieve the minimum MSE of target response matrix and improve the detection probability. To unify performance metrics of sensing and communication, this paper proposes a uplink ISAC design with the objective of maximizing the weighted sum of the normalized sensing rate and the normalized communication rate. In order to reduce the influence of mutual interference, it is required to make a joint optimization of the sensing transmit waveform at the BS transmitter and the communication receive beamforming at the BS receiver.

Nevertheless, it is not a trivial task to design the joint scheme in the presence of mutual interference by using traditional optimization methods due to ultra-complicated expressions, coupled variables and high-dimensional computational complexity. For example, some approximate methods and the alternating direction method of multipliers (ADMM) algorithm was applied to iteratively obtain a sub-optimal solution for joint waveform and beamforming design in the co-existence of multi-input multi-output (MIMO) communication and MIMO radar systems \cite{C19}. The author in \cite{C20} employed an alternating optimization (AO)-based algorithm to design waveform and passive beamforming for reflected intelligence surface (RIS)-aided ISAC system.  Actually, it is seen that the traditional optimization methods for joint waveform and beamforming design mainly rely on some approximate methods to address the nonconvexity of the optimization problem and problem decomposition to decouple the variables. This will undoubtedly lead to a high complexity due to the numerous algorithm iterations and the high dimensional matrix operations, even may not achieve the feasible performance or not meet the real-time requirements of ISAC systems. Moreover, not all complex problems can be solved by traditional optimization methods. To address theses issues, deep learning (DL) offers a novel approach for joint optimization design, because it can transfer complex computational tasks to the offline training phase with rich training samples \cite{DL1}, \cite{DL2}.  Based on this significant benefit, DL technique has been used to solve many classical wireless communication problems, such as channel estimation and signal detection \cite{C21},  beamforming design \cite{C22} and resource allocation \cite{C23}. Motivated by it, this paper is dedicated to provide a novel DL-based joint waveform and beamforming design framework for uplink ISAC to reduce the mutual interference as well as achieve the desired ISAC performance.  The contributions of this paper are three-fold:

\begin{enumerate}

\item We present a general design framework for non-orthogonal uplink ISAC system, where a dual-function BS is deployed to sense nearby targets and serve multiple CUs simultaneously.

\item We propose a joint sensing transmit waveform and communication receive beamforming design to mitigate the mutual interference for implementing sensing information extraction and communication signal decoding at the same time, which is formulated as the weighted sum of the normalized sensing rate and the normalized communication rate maximization problem.

\item We make some equivalent problem transformation to reduce the design complexity, and then design a customized deep neural network (DNN) structure called ``ISACNN" with unsupervised learning according to the characteristics of non-orthogonal uplink ISAC system.

\end{enumerate}
The rest of this paper is outlined as follows. Section II introduces uplink ISAC systems and defines the performance metrics of sensing and communication.  Then, Section III formulates the joint optimization design problem and makes equivalent problem transformation. Next, Section IV provides a DL-based scheme and Section V gives the numerical simulation to validate the effectiveness of the proposed DL-based scheme. Finally, Section VI concludes the paper.

\emph{Notations}: We use bold upper (lower) letters to denote matrices (column vectors), $(\cdot)^H$ to denote conjugate transpose,  $\|\cdot\|_1$ to denote the $\ell_1$-norm of a vector, $\|\cdot\|$ to denote the $\ell_2$-norm of a vector or the $F$-norm of a matrix, $|\cdot|$ to denote the absolute value of a scalar or the determinant of a matrix, $[\cdot]^{\Downarrow}$ to denote the descending operation on the elements of a vector,  $\mathbb{E}\{\cdot\}$ to denote the expectation, ${{\mathbb{C}}^{M\times N}}$ to denote the set of $M$-by-$N$ dimensional complex matrix, ${{\mathbb{R}}^{m\times n}}$ to denote the set of $m$-by-$n$ dimensional real matrix, $\otimes$ to denote the Kronecker product, $\text{vec}(\cdot)$ to denote the vectorization of matrix, $\text{diag}(\mathbf{x})$ to denote a diagonal matrix with the diagonal elements being vector $\mathbf{x}$, $I(\mathbf{A|B;C})$ to denote the MI between $\mathbf{A}$ and $\mathbf{B}$ conditioned on $\mathbf{C}$, and $\mathcal{CN}(\mu,\sigma^2)$ to denote the circularly symmetric complex Gaussian (CSCG) distribution with mean $\mu$ and variance $\sigma^2$.

\section{System Model}
\begin{figure}[h] \centering
\includegraphics [width=0.5\textwidth] {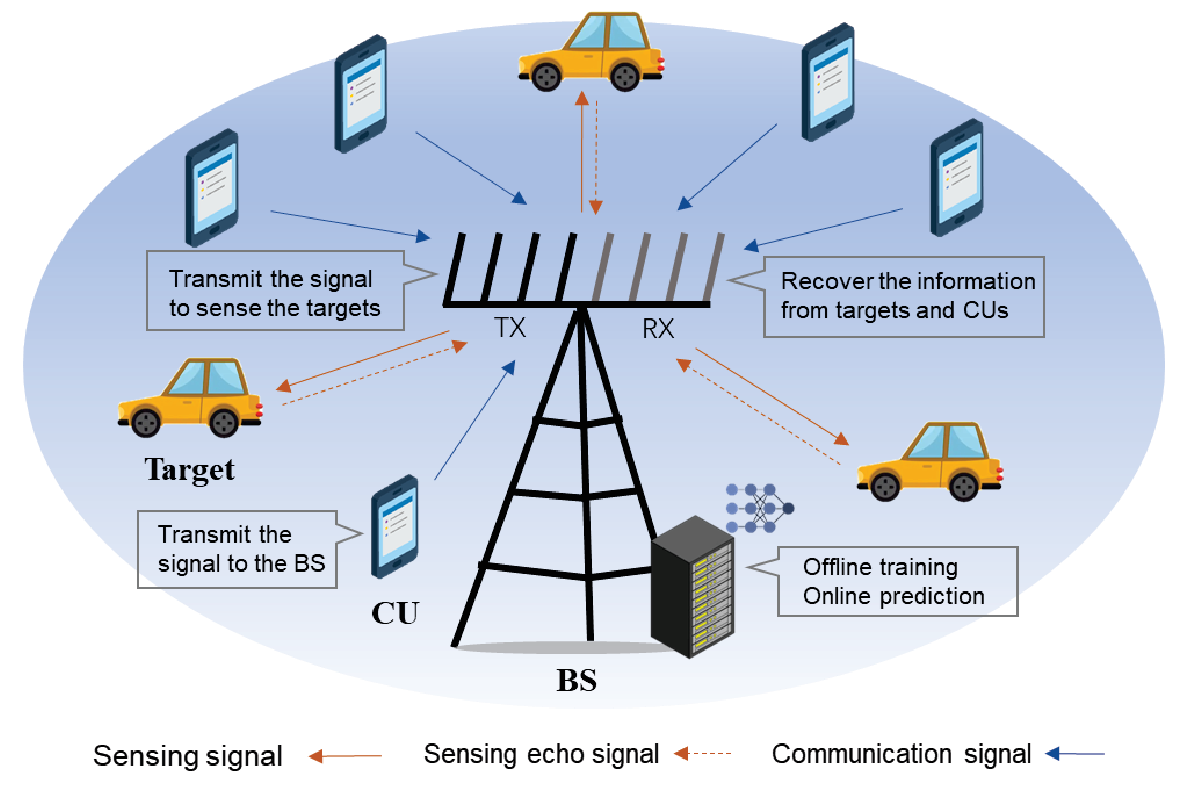}
\caption {An uplink ISAC system.}
\label{Fig1}
\end{figure}

\begin{figure*} \centering
\includegraphics [width=0.9\textwidth] {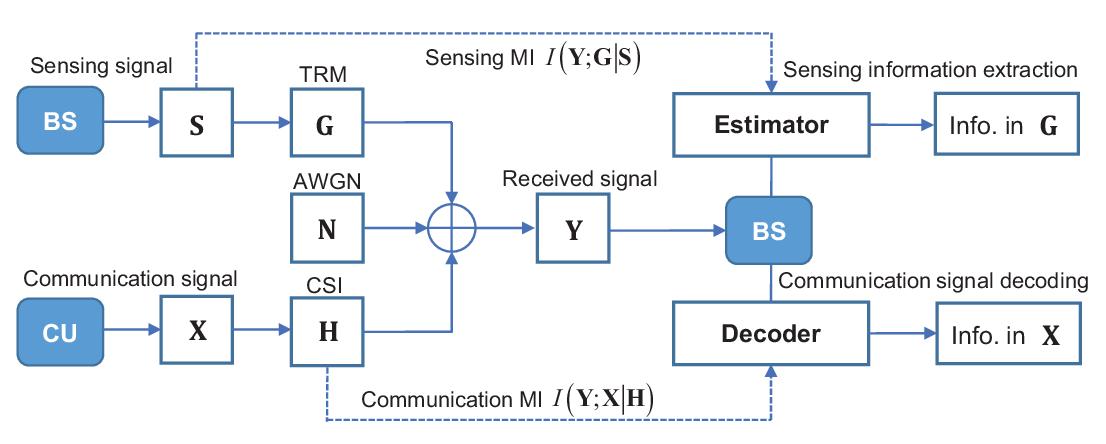}
\caption {System flowchart, where ``Info." denotes information.}
\label{Fig2}
\end{figure*}
As shown in Fig. \ref{Fig1}, this ISAC system deploys a dual-function BS with $N_t$ transmit antennas and $N_r$ receive antennas, serving $K$ single-antenna CUs and sensing nearby several targets at the same time. Specifically, the dual-function BS first broadcasts the sensing signal $\mathbf{S}=[\mathbf{s}_1^T,\ldots,\mathbf{s}_L^T]\in\mathbb{C}^{L\times N_t}$ for nearby environmental sensing, where $L$ is the length of sensing signal waveform with $L>N_t$. Then, the dual-function BS receives the communication signal from the CUs via uplink and the echo signals reflected from targets simultaneously. The received signal $\mathbf{Y}=[\mathbf{y}_1^T,\ldots,\mathbf{y}_L^T]\in\mathbb{C}^{L\times N_r}$ at the BS can be expressed as
\begin{equation} \label{received_Y}
 	\mathbf{Y}=\mathbf{S} \mathbf{G}+\mathbf{XPH}^T+\mathbf{N},
\end{equation}
where $\mathbf{H}=[\mathbf{h}_1,\ldots,\mathbf{h}_K]\in\mathbb{C}^{N_r\times K}$, $\mathbf{P}=\text{diag}[\sqrt{p_1},\ldots,\sqrt{p_K}]\in\mathbb{R}^{K\times K}$ and $\mathbf{X}=[\mathbf{x}_1,\ldots,\mathbf{x}_K]\in\mathbb{C}^{L\times K}$ with $\mathbf{h}_k\in\mathbb{C}^{N_r\times 1}$ being channel state information (CSI) between the BS and the $k$-th CU, $p_k$ and $\mathbf{x}_k\in\mathbb{C}^{L\times 1}$ being the transmit power and transmit signal of the $k$-th CU, respectively. Moreover, $\mathbf{N}=[\mathbf{n}_1,\ldots,\mathbf{n}_{L}]\in\mathbb{C}^{N_r\times L}$ is the additive white Gaussian noise (AWGN) with ${{\mathbf{n}}_{l}}\sim \mathcal{CN}(0,\sigma _{n}^{2})$, and  $\mathbf{G}\in\mathbb{C}^{N_t\times N_r}$ is the target response matrix (TRM) that needs to be sensed, which can be expressed as
\begin{equation}
\mathbf{G}=\sum_{i} \beta_{i} \mathbf{a}\left(\theta_{i}\right) \mathbf{b}^{T}\left(\theta_{i}\right),
\end{equation}
where $\beta_i$ and $\theta_i$ are the reflection coefficient and the direction of arrival (DoA) for the $i$-th target, respectively.   $\mathbf{a}\left(\theta_i\right) $ and $\mathbf{b}\left(\theta_i\right) $ are associated transmit and receive array steering vectors, respectively.
Finally, the BS sends the received signal into the estimator and the decoder for sensing information extraction and communication signal decoding, respectively. The detailed system flowchart is summarized in Fig. \ref{Fig2}.

\begin{remark}
  For the widely distributed antennas at the BS, the difference of column correlations of the TRM $\mathbf{G}$ can be ignored. In this case, we assume that the columns of the TRM $\mathbf{G}$ have the identical correlations for ease of analysis, i.e., ${{\mathbf{R}}_{T}}=\mathbb{E}\left\{ {{\mathbf{g}}_{i}}\mathbf{g}_{i}^{H} \right\},i=1,\ldots ,{{N}_{r}}$, where $\mathbf{g}_{i}\in\mathbb{C}^{N_t\times 1}$ denotes the $i$-th column of the TRM $\mathbf{G}$ \cite{C18}.
\end{remark}

\subsection{Sensing Information Extraction}
After receiving the signal, the BS recovers the TRM $\mathbf{G}$ for nearby targets. Through extracting sensing information from $\mathbf{G}$, the detailed environmental parameters can be obtained, such as the reflection coefficient and the DoA of each target. Considering the MSE minimization of the TRM $\mathbf{G}$ estimation is equivalent to the sensing rate maximization \cite{C13}, \cite{C18}, we focus on the maximization of sensing rate.  To derive the sensing rate, we first give the following useful lemma:
\begin{lemma}\label{lemmavec}
  For arbitrary matrice $\mathbf{A}\in\mathbb{C}^{m\times n}$, $\mathbf{B}\in\mathbb{C}^{n\times p}$ and $\mathbf{C}\in\mathbb{C}^{p\times q}$, we have $\text{vec}\left( \mathbf{ABC} \right)=({{\mathbf{C}}^{T}}\otimes \mathbf{A})\text{vec}\left( \mathbf{B} \right)$.
\end{lemma}
\begin{IEEEproof}
Please refer to Appendix \ref{proof1}.
\end{IEEEproof}
Based on Lemma \ref{lemmavec}, we vectorize the received signal $\mathbf{Y}$ in (\ref{received_Y}), and have
\begin{equation}
  \mathbf{\tilde{y}}=\tilde{\mathbf{S}} \mathbf{\tilde{g}}+\left( \mathbf{H} \mathbf{P}^{T}\otimes \mathbf{I}_{L} \right)\text{vec}(\mathbf{X})+\mathbf{\tilde{n}},\label{vec_Y}
\end{equation}
where  $\mathbf{\tilde{y}}=\text{vec}(\mathbf{Y})\in\mathbb{C}^{LN_r\times 1}$, $\tilde{\mathbf{S}}= \mathbf{I}_{N_{r}}\otimes \mathbf{S}\in\mathbb{C}^{N_rL\times N_rN_t}$, $\mathbf{\tilde{g}}=\text{vec}(\mathbf{G})\in\mathbb{C}^{N_tN_r\times 1}$ and $\mathbf{\tilde{n}}=\text{vec}(\mathbf{N})\in\mathbb{C}^{LN_r\times 1}$. Then, the sensing MI between the vectorized TRM $\mathbf{\tilde{g}}$ and the vectorized receive signal $\mathbf{\tilde{y}}$ with known waveform $\widetilde{\mathbf{S}}$ can be computed as
\begin{eqnarray}
	\begin{aligned}
		I\left(\mathbf{\tilde{y}}; \mathbf{\tilde{g}}\mid  \tilde{\mathbf{S}}\right)&=h\left(\mathbf{\tilde{y}} \mid \tilde{\mathbf{S}}\right)-h\left(\mathbf{\tilde{y}} \mid \mathbf{\tilde{g}}, \tilde{\mathbf{S}}\right) \\
        &={{\log }_{2}}\left| \widetilde{\mathbf{S}}{{\mathbf{R}}_{g}}{{\widetilde{\mathbf{S}}}^{H}}+{{\mathbf{R}}_{I}} \right|-{{\log }_{2}}\left| {{\mathbf{R}}_{I}} \right|\\
		&=\log_2 \left|\mathbf{I}_{L N_{r}}+\mathbf{R}_{I}^{-1} \tilde{\mathbf{S}} \mathbf{R}_{g}   \tilde{\mathbf{S}}^{H}\right|,
	\end{aligned}
\end{eqnarray}
where $h\left( \mathbf{y}|\mathbf{x} \right)=\int{\varrho\left( \mathbf{y}|\mathbf{x} \right){{\log }_{2}}}\varrho\left( \mathbf{y}|\mathbf{x} \right)\text{d}\mathbf{y}$ is the conditional differential entropy of $\mathbf{y}$ for a given $\mathbf{x}$ and $\varrho\left( \mathbf{y}|\mathbf{x} \right)$ denotes the conditional probability density function of $\mathbf{y}$ for a given $\mathbf{x}$. $\mathbf{R}_{I}=\left(\mathbf{H} \mathbf{\tilde{P}}  \mathbf{H}^{H}\right)\otimes\mathbf{I}_{L} +\sigma_{n}^{2} \mathbf{I}_{L N_{r}}$ with $\mathbf{\tilde{P}}=\text{diag}[{p_1},\ldots,{p_K}]\in\mathbb{R}^{K\times K}$ represents the covariance matrix of communication interference plus noise and  $\mathbf{R}_{g}=\mathbf{I}_{N_r} \otimes \mathbf{R}_{T}$ denotes the covariance matrix of the vectorized TRM $\mathbf{\tilde{g}}$. As a result, the sensing rate can be expressed as
\begin{equation} \label{Rs}
  \mathcal{R}_s(\mathbf{S})=I\left(\mathbf{\tilde{y}}; \mathbf{\tilde{g}}\mid  \tilde{\mathbf{S}}\right)/L.
\end{equation}

\subsection{Communication Signal Decoding}
Meanwhile, the decoder at the BS conducts communication signal decoding. As a common indicator, the achievable communication rate is selected as communication performance metric for evaluating communication quality.  During the $l$-th time slot, the received signal at the BS is given by
\begin{equation}
\mathbf{y}_{l}^{T}=\sum\limits_{k=1}^{K}{\sqrt{{{p}_{k}}}}{{x}_{k,l}}\mathbf{h}_{k}^{T}+\mathbf{s}_{l}^{T}\mathbf{G}+\mathbf{n}_{l}^{T},
\end{equation}
where ${x}_{k,l}$ denotes the transmit signal from the $k$-th CU at the $l$-th time slot. To reduce the interference, receive beamforming is employed at the BS for the communication signal from CUs. Thus, the decoded communication signal related to the $k$-th CU at the $l$-th time slot can be expressed as
\begin{equation}
	\hat{{x}}_{k,l} = \sqrt{p_{k}} {x}_{k,l} \mathbf{h}_{k}^{T}\mathbf{w}_{k}+\sum_{i \neq k}^{K} \sqrt{p_{i}} {x}_{i,l} \mathbf{h}_{i}^{T}\mathbf{w}_{k} \nonumber +\mathbf{s}_l^T\mathbf{G}\mathbf{w}_{k} + \mathbf{n}_l^T\mathbf{w}_{k},
\end{equation}
where $\mathbf{w}_{k}\in\mathbb{R}^{N_r\times 1}$ is the receive beamforming for the $k$-th CU. As a result, the corresponding receive signal-to-interference-plus-noise-ratio (SINR) is given by
\begin{equation}
	\gamma_{k}=\frac{p_{k}\left| \mathbf{h}_{k}^T\mathbf{w}_{k}\right|^{2}}{\sum_{i \neq k}^{K} p_{i}\left|\mathbf{h}_{i}^T\mathbf{w}_{k}\right|^{2}+(\left\| \mathbf{S} \mathbf{R}_{T} \mathbf{S}^{H}\right\| /L +\sigma_n^{2})\left\| \mathbf{w}_k\right\|^2}, \forall k, \label{sinr}
\end{equation}
and the achievable communication rate of CUs can be expressed as
\begin{equation}\label{Rc}
	\mathcal{R}_c(\mathbf{w}_k,\mathbf{S})=\frac{1}{K}\sum_{k=1}^{K}\log_{2}(1+\gamma_{k}).
\end{equation}

As is seen from (\ref{Rs}) and (\ref{Rc}),  the sensing rate and the communication rate are jointly affected by sensing transmit waveform $\mathbf{S}$ and the communication receive beamforming $\mathbf{w}_k, \forall k$. Therefore, it makes sense to design a joint waveform and beamforming scheme to improve the overall performance of ISAC system.

\section{Problem Formulation and Transformation}
In this paper, we aim to optimize both sensing and  communication performance under the same resources in ISAC system. On the one hand, maximizing the sensing rate is equal to minimizing the sensing MSE, and thus sensing rate maximization is selected as the objective of sensing function.  On the other hand, maximizing average communication sum-rate can improve the overall throughput by appropriately allocating resources to different users under varying channel conditions. In order to balance communication and sensing, it is more appropriate to choose the maximization of average sum-rate as the objective of communication function.   Based on the principle of multi-objective optimization, we take the weighted sum of the normalized sensing rate and the normalized communication rate as the ultimate optimization objective function.  In particular, the joint waveform and beamforming design is formulated as the following multi-objective optimization problem (MOOP).
\begin{subequations} \label{OP1}
\begin{eqnarray}
\!\!\!\!\!\!\!\!\!\!\!\!\!\!\!\!\!\!\!\!\!\!\!\!\text{\emph{S\&C-MOOP:}}&&\nonumber\\
 \max_{\mathbf{S},\mathbf{w}_k} && \frac{\alpha}{M_s}\mathcal{R}_s(\mathbf{S}) +\frac{(1-\alpha)}{M_c}\mathcal{R}_c(\mathbf{w}_{k},\mathbf{S}) \label{OP1obj}\\
\text { s.t. } && \text{tr}\left(\mathbf{S} \mathbf{S}^{H}\right) \leq P_{s}, \label{OP1st1}
\end{eqnarray}	
\end{subequations}
where $P_s$ is the transmit power budget for sensing signal and  $\alpha$ is the weight to adjust the system preferences between functions of sensing and communication.  Moreover, $M_s$ and $M_c$ are the maximum sensing rate and the maximum communication rate, which are used to make a normalization to respectively scale $\mathcal{R}_s$ and $\mathcal{R}_c$ proportionally and bring them within a specific range, i.e., [0,1], ensuring a balanced impact of each objective function throughout the optimization process.   Such that, the range of weight $\alpha$ can also be simplified to $[0,1]$, where $\alpha$ and $(1-\alpha)$ denote the weights for sensing function and communication function, respectively. Obviously, since the objective function in (\ref{OP1obj}) is computationally difficult and non-convex, it is very tricky to solve problem (\ref{OP1}). In this context, it is necessary to transform the optimization problem for reducing the design complexity. In particular, we first divide this MOOP (\ref{OP1}) into two single-objective optimization problems (SOOPs), and then address the problems of sensing rate maximization and communication rate maximization, respectively. Finally, we recombine the two transformed SOOPs and obtain an equivalent but more manageable MOOP. In the following, we process the sensing SOOP and communication SOOP, respectively.

\begin{remark}
For flexibility of different application scenarios, we do not set QoS constraints, e.g., sensing MSE (or CRB) and communication SINR (or rate), in the proposed MOOP. Specifically, the formulated MOOP is the weighted combination of sensing MOOP and communication SOOP.   Meanwhile, sensing SOOP and communication SOOP are also special examples of MOOP. For instance, by setting function weight $\alpha=1$, the MOOP  can be reduced to a sensing SOOP. In this scenario, there should be no communication QoS constraint. Similarly, the sensing QoS constraint should not exist when $\alpha=0$ since the MOOP is simplified as a communication SOOP. Although such optimization cannot guarantee the threshold of sensing accuracy and non-zero rate for all CUs,  it can achieve a certain level of QoS for sensing and communication by changing the function weight.  It is noted that based on the requirements of the application scenario, future research can also explore other forms of joint optimization problems, such as the joint problem with minimizing the sensing MSE and maximizing the minimum communication rate.
\end{remark}

\subsection{Sensing Rate Maximization}
The SOOP for sensing rate maximization is given by
\begin{eqnarray}
\!\!\!\!\!\!\!\!\!\!\!\!\!\!\!\!\!\!\!\!\!\!\!\!\text{\emph{S-SOOP:}}\ \ \ \ \max_{\mathbf{S}} && \mathcal{R}_s(\mathbf{S})  \label{OP2obj},  \\
\text { s.t. } && (\ref{OP1st1}). \nonumber
\end{eqnarray}	
It is known from $\mathcal{R}_s(\mathbf{S})$ in (\ref{Rs}) that the sensing rate maximization is equivalent to the sensing MI maximization. Thus, we only need to handle the sensing MI, which can be rewritten as
\begin{eqnarray}
		I\left(\mathbf{\tilde{y}}; \mathbf{\tilde{g}}\mid  \tilde{\mathbf{S}}\right)&=&\log_2\left|\mathbf{I}_{L N_{r}}+\mathbf{R}_{I}^{-1} \tilde{\mathbf{S}} \mathbf{R}_{g}   \tilde{\mathbf{S}}^{H}\right| \nonumber \\
		&=&\log_2\left|\mathbf{I}_{L N_{r}}+(\mathbf{R}_{H}\otimes \mathbf{I}_L) (\mathbf{I}_{N_r} \otimes {\mathbf{S}}\mathbf{R}_{T}   {\mathbf{S}^H} )\right| \nonumber \\
		&=&\log_2\left|\mathbf{I}_{L N_{r}}+\mathbf{R}_{H}\otimes {\mathbf{S}} \mathbf{R}_{T}   {\mathbf{S}^{H}}\right|,
\end{eqnarray}
where $\mathbf{R}_{H}=(\mathbf{H} \mathbf{\tilde{P}}  \mathbf{H}^{H}+\sigma_n^2 \mathbf{I}_{N_r})^{-1}$.  In order to reduce the computational dimension, let the singular value decomposition (SVD) of $\mathbf{S}$ be $\mathbf{S}=\mathbf{U}_{s} \boldsymbol{\Sigma}_{s} \mathbf{V}_{s}^{H}$, where $\left.\boldsymbol{\Sigma}_{s}=\left[\left(\boldsymbol{\Sigma}_{s}^{\downarrow}\right)^{1 / 2}, \mathbf{0}_{N_t \times\left(L-N_{t}\right)}\right)\right]^{T}$ is the diagonal matrix of $\mathbf{S}$ with $\boldsymbol{\Sigma}_{s}^{\downarrow}=\text{diag}\left(\left[\sigma_{s, 1}, \ldots, \sigma_{s, N_{t}}\right]^{T}\right)$, $\sqrt{\sigma_{s,k}}$ is  the  $k$-th singular value of $\mathbf{S}$ with $ \sigma_{s, 1} \geq \sigma_{s, 2} \geq \cdots \geq \sigma_{s, N_{t}}$, $\mathbf{U}_{s} $ and $\mathbf{V}_{s} $ are left and right singular vectors of $\mathbf{S}$, respectively.
 Meanwhile, the eigenvalue decomposition (EVD) is conducted on $\mathbf{R}_T$ and $\mathbf{R}_H$. Specifically, $\mathbf{R}_T$ can be expressed as $\mathbf{R}_T=\mathbf{U}_{T} \boldsymbol{\Sigma}_{T} \mathbf{U}_{T}^{H}$, where $\boldsymbol{\Sigma}_{T}=\text{diag}\left(\left[\sigma_{t, 1}, \ldots, \sigma_{t, N_t}\right]^{T}\right)$, ${\sigma_{t,i}}$ is  the  $i$-th eigenvalue of $\mathbf{R}_T$ with  $ \sigma_{t, 1} \geq \sigma_{t, 2} \geq \cdots \geq \sigma_{t, N_{t}}$ and $\mathbf{U}_{T}$ is the corresponding eigenvector. $\mathbf{R}_H$ is decomposed as $\mathbf{R}_H=\mathbf{U}_{h} \boldsymbol{\Sigma}_{h} \mathbf{U}_{h}^{-1}$, where  $\boldsymbol{\Sigma}_{h}=\text{diag}\left(\left[\sigma_{h, 1}, \ldots, \sigma_{h, N_r}\right]^{T}\right)$, ${\sigma_{h,j}}$ is the  $j$-th eigenvalue of $\mathbf{R}_H$ with $ \sigma_{h, 1} \geq \sigma_{h, 2} \geq \cdots \geq \sigma_{t, N_{r}}$ and $\mathbf{U}_{h}$ is the corresponding eigenvector. Thus, the sensing MI can be transformed as
 \begin{eqnarray}\label{MI2}
   &\!\!\!\!&\!\!\!\!I\left(\mathbf{\tilde{y}}; \mathbf{\tilde{g}}\mid  \tilde{\mathbf{S}}\right)=\\ \nonumber
   &\!\!\!\!&\!\!\!\!{{\log }_{2}}\left| {{\mathbf{I}}_{L{{N}_{r}}}}+{{\mathbf{U}}_{h}}{{\mathbf{\Sigma }}_{h}}\mathbf{U}_{h}^{-1}\otimes {{\mathbf{U}}_{s}}{{\mathbf{\Sigma }}_{s}}\mathbf{V}_{s}^{H}{{\mathbf{U}}_{T}}{{\mathbf{\Sigma }}_{T}}\mathbf{U}_{T}^{H}{{\mathbf{V}}_{s}}\mathbf{\Sigma }_{s}^{H}\mathbf{U}_{s}^{H} \right|.
 \end{eqnarray}
To further process the sensing MI in (\ref{MI2}), we propose the following lemma:

\begin{lemma} \label{lemma2}
For square matrices $\mathbf{A}\in \mathbb{C}^{m \times m}, \mathbf{B} \in \mathbb{C}^{m \times m}$, $\mathbf{C} \in \mathbb{C}^{n \times n}$ and $\mathbf{D} \in \mathbb{C}^{n \times n}$, we have $\left|\mathbf{I}_{mn}+\mathbf{AB}\otimes \mathbf{CD} \right| = \left|\mathbf{I}_{mn}+\mathbf{BA}\otimes \mathbf{DC} \right| $.
\end{lemma}
\begin{IEEEproof}
Please refer to Appendix \ref{proof2}.
\end{IEEEproof}
Based on Lemma \ref{lemma2}, the sensing MI can be converted as
\begin{eqnarray}
	&&\!\!\!I\left(\mathbf{\tilde{y}}; \mathbf{\tilde{g}}\mid  \tilde{\mathbf{S}}\right)\nonumber\\
&=&{{\log }_{2}}\left| {{\mathbf{I}}_{L{{N}_{r}}}}+{{\mathbf{\Sigma }}_{h}}\otimes {{\mathbf{\Sigma }}_{s}}\left( \mathbf{V}_{s}^{H}{{\mathbf{U}}_{T}} \right){{\mathbf{\Sigma }}_{T}}\left( \mathbf{U}_{T}^{H}{{\mathbf{V}}_{s}} \right)\mathbf{\Sigma }_{s}^{H} \right|  \nonumber\\
	&\leq&\log_2 \left| \mathbf{I}_{L N_{r}}+ \boldsymbol{\Sigma}_{h}\otimes  \boldsymbol{\Lambda}\right| , \label{upper}
\end{eqnarray}
where $\mathbf{\Lambda }={{\mathbf{\Sigma }}_{s}}{{\mathbf{\Sigma }}_{T}}\mathbf{\Sigma }_{s}^{H}=\left[ \begin{matrix}
   \mathbf{\Sigma }_{s}^{\downarrow }{{\mathbf{\Sigma }}_{T}} & \mathbf{0}  \\
   \mathbf{0} & \mathbf{0}  \\
\end{matrix} \right]\in {{\mathbb{R}}^{L\times L}}$. Note that the upper bound in (\ref{upper}) is achieved if and only if $\mathbf{V}_s=\mathbf{U}_T$ \cite{C18}. In this context, the sensing rate can be rewritten as
\begin{equation}
\mathcal{\bar{R}}_s(\bm{\sigma}_{s})=\frac{1}{L}\sum_{i=1}^{N_t}\sum_{j=1}^{N_r}\log_2 (1+\sigma_{t,i}\sigma_{s,i}\sigma_{h,j}),
\end{equation}
where $\bm{\sigma}_{s}=[{\sigma}_{s,1},\ldots,{\sigma}_{s,N_t}]^T$. Meanwhile, the power constraint (\ref{OP1st1}) is replaced by
\begin{equation} \label{stnew}
    \text{tr}\left(\mathbf{S} \mathbf{S}^{H}\right) =\sum_{i=1}^{N_t}{\sigma_{s,i}}=\|\bm{\sigma}_{s}\|_1\leq P_s.
\end{equation}
Thus, the sensing SOOP can be transformed as
\begin{eqnarray}
\!\!\!\!\!\!\!\!\!\!\!\!\!\!\!\!\!\!\!\!\!\!\!\!\text{\emph{T-S-SOOP:}}\ \ \ \ \max_{\bm{\sigma}_s} && \mathcal{\bar{R}}_s(\bm{\sigma}_{s})  \label{OP3obj},  \\
\text { s.t. } && (\ref{stnew}). \nonumber
\end{eqnarray}

\subsection{Communication Rate Maximization}
On the other hand, the SOOP for communication rate maximization can be expressed as
\begin{eqnarray}
\!\!\!\!\!\!\!\!\!\!\!\!\!\!\!\!\!\!\!\!\!\!\!\!\text{\emph{C-SOOP:}}\ \ \ \ \max_{\mathbf{S},\mathbf{w}_k} && \mathcal{R}_c(\mathbf{w}_{k},\mathbf{S}) \label{OP4obj}\\
\text { s.t. } && (\ref{OP1st1}). \nonumber
\end{eqnarray}	
It is found from $\mathcal{R}_c(\mathbf{w}_{k},\mathbf{S})$ in (\ref{Rc}) that the variables $\mathbf{S}$ and $\mathbf{w}_k$ are separable from each other. Thus, we can consider the communication rate maximization in terms of sensing transmit waveform and communication receive beamforming, respectively.
For a given $\mathbf{S}$, the communication rate maximization in terms of $\mathbf{w}_k$ is equivalent to SINR maximization for each CU as follows
\begin{equation}
	\max _{\mathbf{w}_{k}} \gamma_{k}. \label{SINR_MAX}
\end{equation}	
By utilizing $\left\| \mathbf{S} \mathbf{R}_{T} \mathbf{S}^{H}\right\|=\text{tr}(\boldsymbol{\Lambda})=\sum_{i=1}^{N_t}{\sigma_{s, i}\sigma_{t,i}}$, the communication SINR $\gamma_{k}$ in (\ref{SINR_MAX}) can be rewritten as
\begin{equation}\label{SINR_NEW}
 \gamma_{k}(\bm{\sigma}_s)=\frac{p_{k}\left|\mathbf{h}_{k}^T\mathbf{w}_{k}\right|^{2}}{\sum_{i \neq k}^{K} p_{i}\left|\mathbf{h}_{i}^T\mathbf{w}_{k}\right|^{2}+(\sum_{i=1}^{N_t}{\sigma_{s, i}\sigma_{t, i}}/L+\sigma_n^{2})\left\| \mathbf{w}_k\right\| ^2}.
\end{equation}
Then, let the numerator be equal to 1 in (\ref{SINR_NEW}),  problem (\ref{SINR_MAX}) can be transformed as the denominator minimization problem, which can be expressed as
\begin{subequations}
\begin{eqnarray}
	\min _{\mathbf{w}_{k}} &&\mathbf{w}_k^H\mathbf{\tilde{R}}_k\mathbf{w}_k\\
	\text { s.t. } && \sqrt{p_k}\mathbf{w}_k^H\mathbf{h}_k^{*}=1,
\end{eqnarray}	
\end{subequations}
where $\mathbf{\tilde{R}}_k=\sum_{i \neq k}^{K} p_{i} \mathbf{h}_{i}^{*}\mathbf{h}_{i}^T+(\sum_{i=1}^{N_t}{\sigma_{s, i}\sigma_{t, i}}/L+\sigma_n^{2})\mathbf{I}_{N_r}$. To solve this problem, we can adopt the commonly used Lagrange multiplier method. Specifically, we first construct the Lagrange function of $\mathbf{w}_k$ as
\begin{equation} \mathcal{L}_c(\mathbf{w}_k)=\mathbf{w}_k^H\mathbf{\tilde{R}}_k\mathbf{w}_k+\lambda(\sqrt{p_k}\mathbf{w}_k^H\mathbf{h}_k^{*}-1),
\end{equation}
where $\lambda>0$ is the Lagrange multiplier. Then, let the first derivative of $\mathcal{L}_c(\mathbf{w}_k)$ be $\mathbf{0}$, we have
\begin{equation}
  \frac{\partial \mathcal{L}_c(\mathbf{w}_k)}{\partial \mathbf{w}_k}=2 \mathbf{\tilde{R}}_k \mathbf{w}_k+\lambda \sqrt{p_k}{\mathbf{h}}_k^*=\mathbf{0}.
\end{equation}
Thus, $\mathbf{w}_k$ can be expressed as
\begin{equation}
	\mathbf{w}_k=-\frac{\sqrt{p}_k}{2}\lambda \mathbf{\tilde{R}}_k^{-1}\mathbf{h}_k^*. \label{receiver}
\end{equation}
By left-multiplying $\sqrt{p_k}\mathbf{h}_k^T$ on (\ref{receiver}), we can obtain
\begin{equation} \lambda=-\frac{2}{{p_k}\mathbf{h}_k^T\mathbf{\tilde{R}}_k^{-1}\mathbf{h}_k^*}. \label{lambda1}
\end{equation}
Taking (\ref{lambda1}) into (\ref{receiver}), the optimal communication receive beam for communication rate maximization is given by
\begin{equation} \label{r_opt}
   \mathbf{w}_k^{\text{opt}}=\frac{1}{\sqrt{p_k}\mathbf{h}_k^T\mathbf{\tilde{R}}_k^{-1}\mathbf{h}_k^*}\mathbf{\tilde{R}}_k^{-1}\mathbf{h}_k^*.
\end{equation}
It is found from (\ref{r_opt}) that the optimal communication receive beam $\mathbf{w}_k^{\text{opt}}$ can be viewed as the function of the sensing transmit waveform $\mathbf{S}$. As a result, the communication rate can be rewritten as
\begin{equation}
   \mathcal{\bar{R}}_c(\bm{\sigma}_{s})=\frac{1}{K}\sum_{k=1}^{K}\log_{2}(1+\gamma_k^{\text{opt}}), \label{newRc}
\end{equation}
where $\gamma_k^{\text{opt}}=p_k\mathbf{h}_k^T\mathbf{\tilde{R}}_k^{-1}\mathbf{h}_k^*$ is the $\text{SINR}$ for the optimal communication receive beam $\mathbf{w}_k^{\text{opt}}$  at the $k$-th CU. Thus, the communication SOOP is transformed as
\begin{eqnarray}
\!\!\!\!\!\!\!\!\!\!\!\!\!\!\!\!\!\!\!\!\!\!\!\!\text{\emph{T-C-SOOP:}}\ \ \ \ \max_{\bm{\sigma}_s} && \mathcal{\bar{R}}_c(\bm{\sigma}_{s})  \label{OP5obj},  \\
\text { s.t. } && (\ref{stnew}). \nonumber
\end{eqnarray}	

\begin{figure*} \centering
\includegraphics [width=1.0\textwidth] {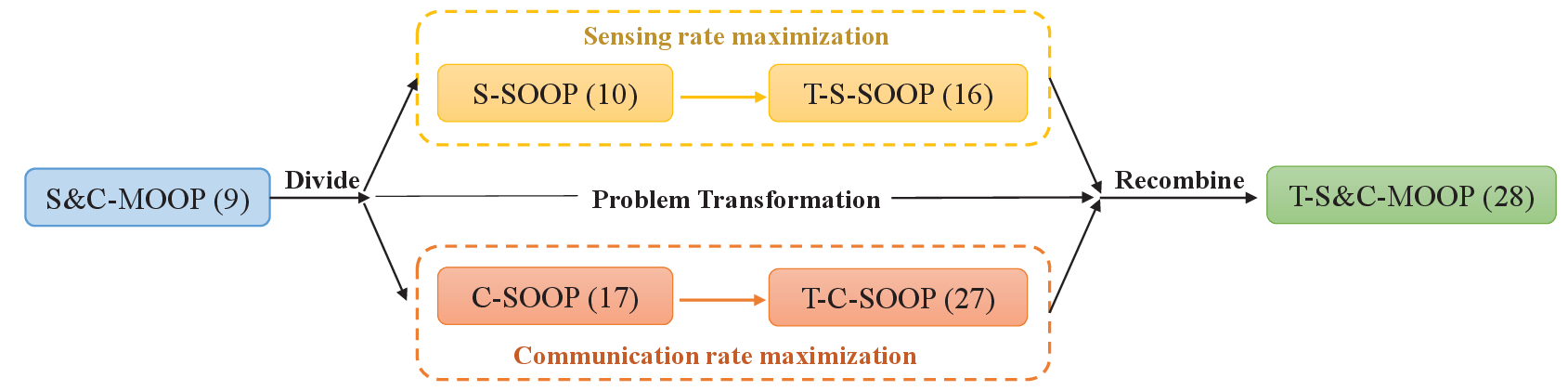}
\caption {Process of problem transformation.}
\label{Fig_trans}
\end{figure*}
\subsection{Problem Transformation}

By recombining SOOPs (\ref{OP3obj}) and (\ref{OP5obj}), problem (\ref{OP1}) can be equivalently converted as
\begin{eqnarray}
\text{\emph{T-S\&C-SOOP:}} && \nonumber \\
\ \ \ \ 	\max _{\bm{\sigma}_{s}}&&\frac{\alpha}{M_s}\mathcal{\bar{R}}_s(\bm{\sigma}_{s})+\frac{(1-\alpha)}{M_c}\mathcal{\bar{R}}_c(\bm{\sigma}_{s}) \label{op6obj}\\
	\text { s.t. } && (\ref{stnew}), \nonumber
\end{eqnarray}	
where the detailed derivations of the maximum sensing rate ${M_s}$ and the maximum communication rate ${M_c}$ are shown in Appendix B. For more intuitive, the problem transformation process is presented in Fig. \ref{Fig_trans}.

Although the computational dimension and design complexity are greatly reduced compared to the original problem (\ref{OP1}), the transformed problem (\ref{op6obj}) is still very difficult to be solved by traditional optimization approaches, due to the extremely complicated structure of objective function. In this context, we turn to develop an efficient DL-based scheme to address this problem for improving the overall performance of ISAC.

\section{DL-based Scheme Design}
\begin{figure*} \centering
\includegraphics [width=1.0\textwidth] {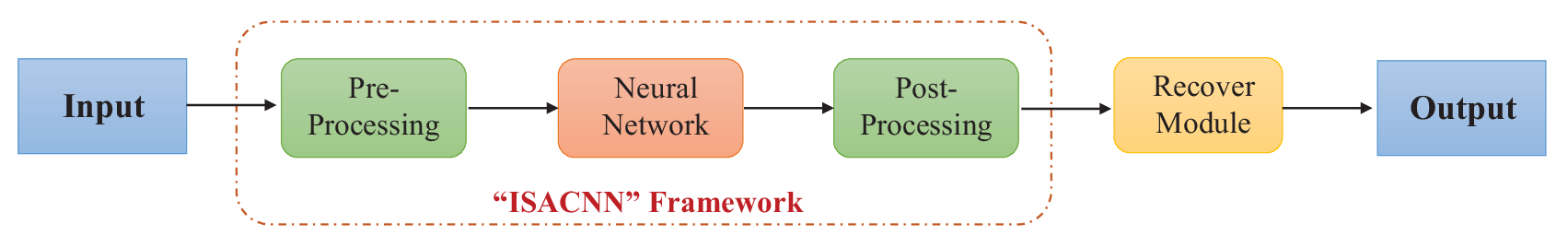}
\caption {The Proposed DL-based scheme for joint waveform and beamforming design.}
\label{Fig_NNFrame}
\end{figure*}
\begin{figure*} \centering
\includegraphics [width=1.0\textwidth] {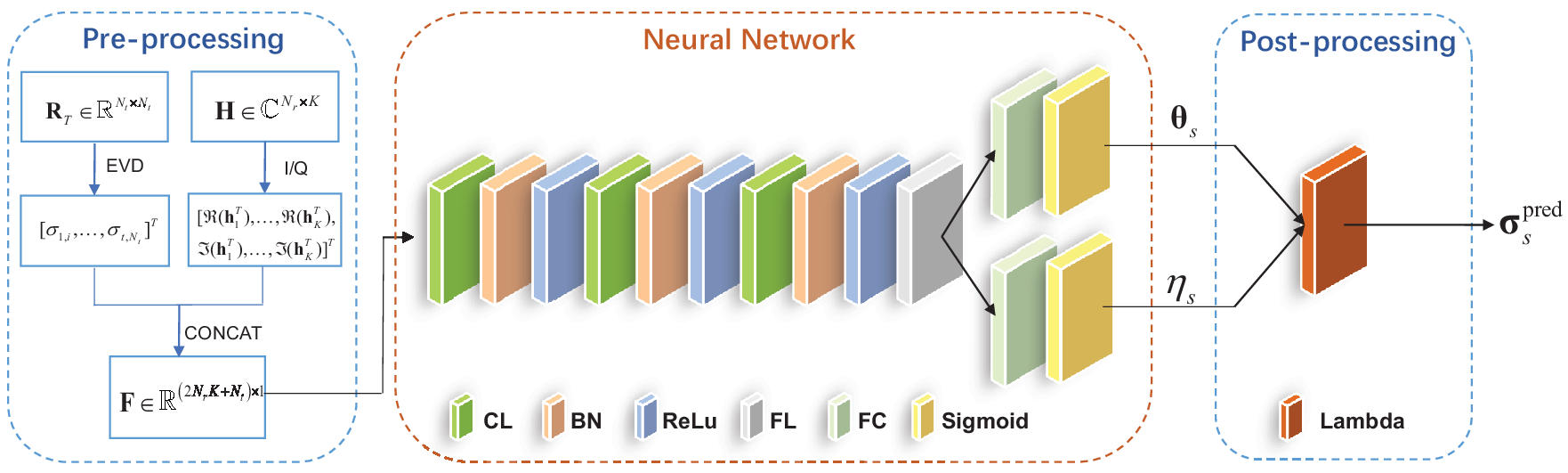}
\caption {Framework of ``ISACNN".}
\label{Fig_NN}
\end{figure*}

To solve the transformed optimization problem (\ref{op6obj}), we propose a DL-based joint design scheme to obtain a feasible solution. It is known that the sensing transmit waveform $\mathbf{S}$ and the communication receive beamforming $\mathbf{w}_k, \forall k$ are mainly affected by the sensing transmit covariance matrix (TCM) $\mathbf{R}_T$ and the communication CSI $\mathbf{H}$. Thus, the input of the proposed DL-based scheme are sensing TCM $\mathbf{R}_T$ and the communication CSI $\mathbf{H}$, while its output are the sensing transit waveform $\mathbf{S}$ and the communication receive beamforming $\mathbf{w}_k$. As shown in Fig. \ref{Fig_NNFrame}, the core of DL-based joint design scheme is the DL network, called ``ISACNN", which mainly includes pre-processing,  network structure and post-precessing, c.f. Fig. \ref{Fig_NN}. Finally, for a given predicted vector from ``ISACNN", the recover module based on the derivation of Section III is used to obtain the desired solutions $\mathbf{S}$ and $\mathbf{w}_k$. In the following, we will give the detail discussion of the proposed DL-based scheme.

\subsection{Pre-processing}
 As mentioned above, the input of the proposed DL-based scheme are sensing TCM $\mathbf{R}_T$ and the communication CSI $\mathbf{H}$.  However, the current DL network can not process the complex input vector. In this context, we adopt the commonly used I/Q transformation to divide the complex channel vector $\mathbf{h}_k$ into in-phase component $\mathfrak{R}(\mathbf{h}_k^T)$ and quadrature component $\mathfrak{I}(\mathbf{h}_k^T)$, where $\mathfrak{R}(\cdot)$ and $\mathfrak{I}(\cdot)$ denote the real part and the imaginary part, respectively. Meanwhile, based on the transformed optimization problem (\ref{op6obj}), the input of sensing TCM can be replaced by its eigenvalue $\sigma_{t,i}, i=1,\ldots,N_t$. Thus, we contact the real part and imaginary part of $\mathbf{h}_k, k=1,\ldots,K$ and the eigenvalue of the sensing TCM $\sigma_{t,i}, i=1,\ldots,N_t$  as a feature input vector $\mathbf{F}\in\mathbb{R}^{(2N_rK+N_t)\times 1}$ for ``ISACNN", i.e.,
\begin{equation}\label{int}
  \mathbf{F}=[\mathfrak{R}(\mathbf{h}_1^T),\ldots,\mathfrak{R}(\mathbf{h}_K^T),\mathfrak{I}(\mathbf{h}_1^T),\ldots,\mathfrak{I}(\mathbf{h}_K^T),\sigma_{1,i},\ldots,\sigma_{t,N_t}]^T.
\end{equation}

 \subsection{Post-processing}
 According to the transformed optimization problem, we select the square of the singular value for the sensing transmit waveform $\sigma_{s,i}, i=1,\ldots,N_t$ as the output of ``ISACNN". In order to ensure that network output vector meets the sensing transmit power constraint (\ref{stnew}), we turn to train two feature parameters. One is a vector $\bm{\theta}_s$, the other is a scale scalar $\eta_s$. Then, the two feature parameters are contacted to send to the Lambda layer for satisfying the constraint. Specifically, the function of the Lambda layer can be realized by the following steps:
 \begin{enumerate}
 \item Separate out the vector $\bm{\theta}_s$ and the scalar $\eta_s$;

 \item Normalize the vector $\bm{\theta}_s$ and sort its elements from largest to smallest, i.e., ${\bm{\theta }_{s}}={{\left[ \frac{{\bm{\theta }_{s}}}{\left\| {\bm{\theta }_{s}} \right\|} \right]}^{\Downarrow }}$;

 \item Multiply the scalar $\eta_s$ by $P_s$, i.e., $\eta_s=\eta_s \times P_s$;

 \item Multiply the $\bm{\theta}_s$ by $\eta_s$ to obtain the output vector, i.e., $\bm{\sigma}_s^{\text{pred}}=\bm{\theta}_s \times \eta_s$.
 \end{enumerate}

Based on the function of Lambda layer, the norm of vector $\bm{\theta}_s$ is always 1 and the range of scalar $\eta_s$ is scaled to $[0, P_s]$. Thus, the output vector $\bm{\sigma}_s^{\text{pred}}=\bm{\theta}_s \times \eta_s$ always meets the constraint, i.e., $\|\bm{\sigma}_{s}^{\text{pred}}\|\leq P_s$.

\subsection{Network Structure}
With the preprocessing and the post-processing, the adopted architecture of the neural network for ``ISACNN" is composed of input layer (feature input vector), convolution (CL) layers, batch-normalization (BN) layers, flatten (FL) layers, fully-connected (FC) layers, activation layers with rectified linear unit (ReLU) or Sigmoid functions, and Lambda layer (output). Specifically, the CL layer is used to extract the features from the input vector, while the FC layer can integrate the feature information from the CL layer. This is because each neuron in the FC layer is fully connected to all the neurons in the layer before it. In our proposed NN, two FC layers with $N_t$ and $1$ neurons are used to predict the vector $\bm{\theta}_s$ and the scalar $\eta_s$, respectively. The FL layer is commonly used in transitions from the CL layer to the FC layer, whose purpose is to "flatten" the input, i.e., turning multi-dimensional inputs into one-dimensional inputs. For the activation functions, the CL layers all use the ReLU function, while the two FC layers both adopt the Sigmoid function, which are mainly determined based on the empirical experiment.  To accelerate the convergence, each CL layer and each FC layer are preceded by a BN layer. In the end, the output layer is imposed by a Lambda layer to scale and sort the training output result for ensuring the constraint.

Unlike traditional supervised learning designs, no training labels are required in our proposed ``ISACNN" framework.  To directly improve the system performance, we adopt the negative objective in (\ref{op6obj}) as the loss function. In this context, the decrease of the training loss exactly corresponds to the increase of the average weighted sum of normalized sensing rate and normalized communication rate. The defined loss function can be expressed as
\begin{equation}\label{loss}
  \text{Loss}=-\frac{1}{{{N}_{s}}}\sum\limits_{q=1}^{{{N}_{s}}}{\left[ \frac{\alpha }{{{M}_{s,q}}}{{{\bar{\mathcal{R}}}}_{s}}(\bm{\sigma }_{s,q}^{\text{pred}})+\frac{(1-\alpha )}{{{M}_{c,q}}}{{{\bar{\mathcal{R}}}}_{c}}(\bm{\sigma }_{s,q}^{\text{pred}}) \right]},
\end{equation}
where $N_s$ is the number of training samples,  $\mathbf{\sigma }_{s,q}$ is the predict output vector with the $q$-th sample, ${M}_{s,q}$ and ${M}_{c,q}$ are the maximum sensing rate and maximum communication rate for the $q$-th sample, respectively.

\subsection{Recover Module}
Finally, the recover module is used to obtain the desired sensing transmit waveform and the communication receive beamforming based on the output vector $\bm{\sigma}_{s}^{\text{pred}}$. In particular, the sensing transmit waveform can be computed by
\begin{equation}\label{ssw}
  \mathbf{S}={{\mathbf{U}}_{s}}{\bm{\Sigma }_s^{\text{pred}}}\mathbf{U}_{T}^{H},
\end{equation}
where ${\mathbf{U}}_{s}$ is any $L$-dimensional unitary matrix and ${{\bm{\Sigma }_s}^{\text{pred}}}=[\text{diag}\left( ({\bm{\sigma}_{s}^{\text{pred}}})^{1/2} \right),\mathbf{0}_{N_t\times(L-N_t)}]^T$. Moreover, the communication receive beamforming is given by
\begin{equation}\label{crb}
  {{\mathbf{w}}_{k}}=\frac{1}{\sqrt{{{p}_{k}}}\mathbf{h}_{k}^{T}{{\left( \bm{\Omega }_{k}^{\text{pred}} \right)}^{-1}}\mathbf{h}_{k}^{*}}{{\left( \bm{\Omega }_{k}^{\text{pred}} \right)}^{-1}}\mathbf{h}_{k}^{*},
\end{equation}
where $\bm{\Omega }_{k}^{\text{pred}}=\sum\limits_{i\ne k}^{K}{{{p}_{i}}}\mathbf{h}_{i}^{*}\mathbf{h}_{i}^{T}+({\bm{\sigma }_{t}^{T}\bm{\sigma }_{s}^{\text{pred}}}/{L}\;+\sigma _{n}^{2}){{\mathbf{I}}_{{{N}_{r}}}}$.

\begin{table*}[t]
\centering
\caption{The real average runtime (ms) of proposed DL scheme}\label{runtime}
\begin{tabular}{|c|c|c|c|c|c|c|c|}
\hline
$N_t$ & 4 & 6 & 8 & 10& 12 & 14 & 16 \\\hline
Time (ms) &0.2330 &0.2360 & 0.2404& 0.2418 &0.2422 &0.2447 &0.2457 \\\hline
$N_r$ & 4 & 6 & 8 & 10 & 12 & 14 & 16 \\\hline
Time (ms) &   0.1413  &  0.1572  &  0.1732  &  0.1943  &  0.2052  &  0.2206  &  0.2504 \\\hline
$K$ & 2 & 3 & 4 & 5 & 6 & 7 & 8 \\\hline
Time (ms)& 0.1653  &  0.1897  &  0.2169  &  0.2504  &  0.2687   & 0.2812  &  0.2934 \\\hline
\end{tabular}
\end{table*}

\subsection{Complexity Analysis}
 Herein, we analyze the computational complexity of the proposed DL-based scheme, which mainly includes pre-processing, prediction and recovery. For pre-processing stage, EVD operation is used to acquire the eigenvalue, whose complexity is $\mathcal{C}_1=\mathcal{O}(N_t^3)$. Then, for prediction stage, based on \cite{complexity1} and \cite{complexity2}, the complexity for the forward calculation of the network can be expressed as $\mathcal{C}_2=\mathcal{O}\left( \sum\limits_{l=1}^{N_L-2}{C_{l}^{F}C_{l}^{K}{{F}_{l-1}}{{F}_{l}}}+\sum\limits_{l=N_L-1}^{N_L}{{{F}_{l-1}}{{F}_{l}}} \right)$, where $N_L$ is the number of layers, $C^F_l,C^K_l,F_{l-1},F_l$ are the area of the feature map for the $l$-th CL layer, the area of the convolution kernel for the $l$-th CL layer, the number of the input channel for the $l$-th layer and the number of the output channel for the $l$-th layer, respectively.   For recovery stage, the complexity of recovering sensing transmit waveform and communication receive beamforming is given by $\mathcal{C}_3=\mathcal{O}(KN_r^3+L^2N_t^2)$. Thus, the complexity for our proposed ``ISACNN" is given by $\mathcal{C}_1+\mathcal{C}_2+\mathcal{C}_3$.  To make it more vivid, we list the real average runtime (ms) of the proposed DL-based scheme in Tab. \ref{runtime}, where the default setting is according to Section V.

\begin{table*}
\small
\centering
\caption{Simulation Parameters }\label{Simulation}
\begin{tabular}{|c|c|}
\hline
Parameters & Values \\ \hline
Number of transmit antennas at the BS& $N_t=16$  \\\hline
Number of receive antennas at the BS& $N_r=16$  \\\hline
Number of CUs& $K=5$  \\\hline
Cell radius & $d_0=200$ m \\\hline
Weight of sensing rate &  $\alpha=0.5$ \\\hline
Length of sensing signal waveform & $L=20$ \\\hline
Maximum sensing transmit SNR at the BS & $\text{SNR}_s=10 $ dB  \\\hline
Maximum communication transmit SNR at the CUs & $\text{SNR}_c=0 $ dB  \\\hline
\end{tabular}
\end{table*}

\begin{figure*}
  \centering
  \includegraphics[width=1\textwidth]{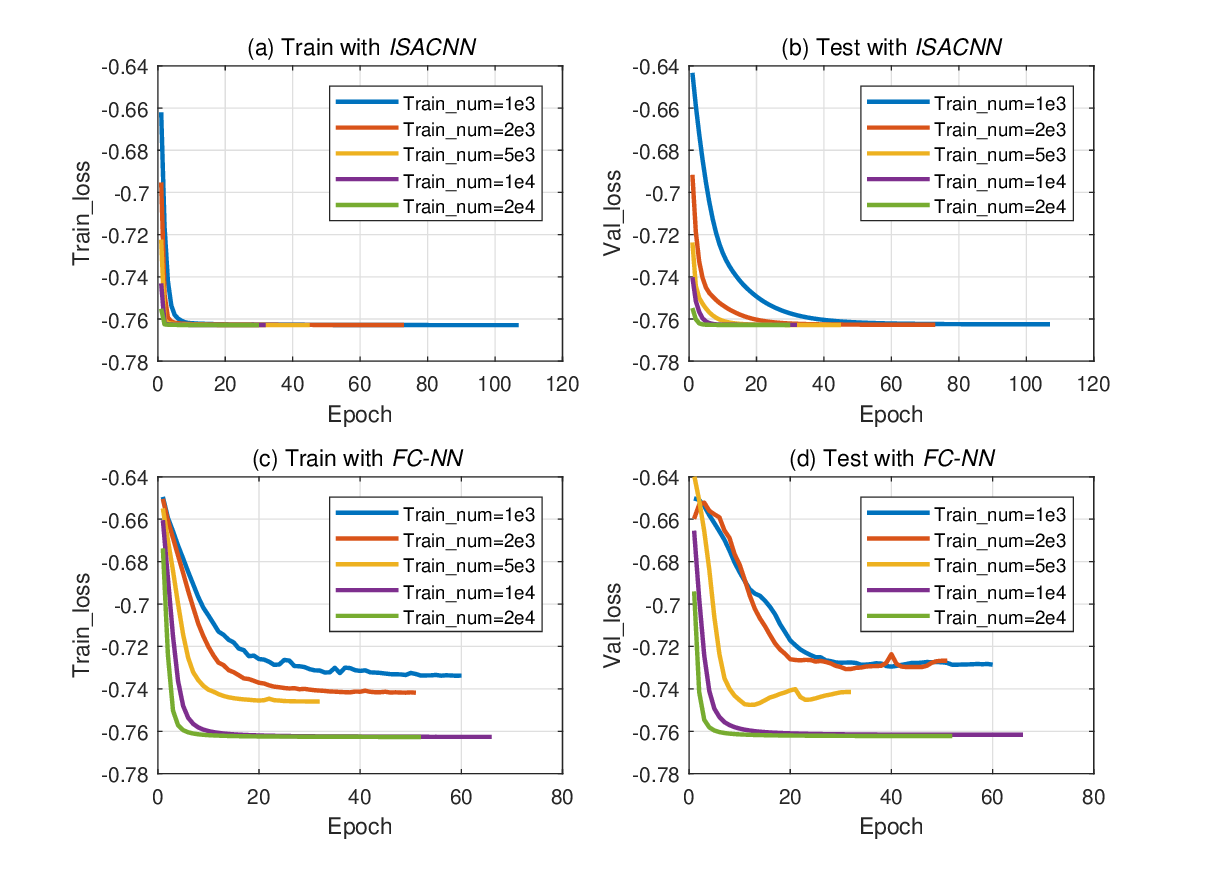}
  \caption{The training and testing performance versus epoch with different NNs.}
  \label{Fig_train}
\end{figure*}
\section{Simulation Results}

In this section, we give some numerical results to validate the effectiveness of the proposed DL-based scheme. For system setting, the main parameters are listed in Tab. \ref{Simulation}\footnote{It is worth pointing out that our proposed model and scheme are applicable to any number of CUs. In practical scenarios, the number of activated CUs accessing the same BS over the same resource block in a time slot is usually limited. Thus, we chose an appropriate number of CUs in simulation experiments to evaluate the system's performance.}.
 It is assumed that all CUs and STs are randomly distributed beyond the cell radius $d_0$ and the pathloss model is $\mathrm{PL}_{\mathrm{dB}}=128.1+37.6\log_{10}(d)$ \cite{pathlossmodel}, where $d$ (km) denotes the distance. The Rayleigh fading is considered as the small-scale fading of communication channels. Specifically, the channel of the $k$-th CU is modeled as $\mathbf{h}_k=\sqrt{{{\xi }_{k}}}\mathbf{\bar{h}_k}$, where $\mathbf{\bar{h}_k} \sim \mathcal{CN}(0, \mathbf{I})$ and ${{\xi }_{k}}$ is the pathloss coefficient from the $k$-th CU to the BS.  For example,  with $d_k$ being the distance between the $k$-th CU to the BS, the related pathloss coefficient  is given by ${{\xi }_{k}}=10^{(-PL_k/10)}=10^{(-12.81-3.76 \log_{10}(d_k))}=10^{(-12.81)}/(d_k)^{3.76}$.  As the distance is equal to the cell radius $d_0$, the pathloss coefficient ${{\xi }_{0}}$ can be computed as ${{\xi }_{0}}=10^{(-12.81)}/(d_0)^{3.76}$.  The sensing TCM $\mathbf{R}_T$ is randomly generated according to \cite{RT}, whose eigenvector is obtained by performing SVD of random Gaussian matrix with i.i.d. entries whereas the eigenvalues are generated by considering i.i.d. uniform (positive) random variables. Here, we use $\text{SNR}_s=10\log_{10}(P_{s}\cdot {{\xi }_{0}}/\sigma_n^2)$ to denote the sensing signal-to-noise ratio (SNR) (in dB) of the BS and  $\text{SNR}_k=10\log_{10}(p_k\cdot {{\xi }_{k}}/\sigma_n^2)$ to denote the communication SNR (in dB) of the CUs, where  all CUs are assumed to have the same $\text{SNR}_c$ for ease of analysis.

For DL training setting, we generate 10000 training samples and set the validation split factor as 0.2 (i.e., 8000 samples for training and 2000 samples for validation) to evaluate the training effect at the offline training phase. Note that through a series of equivalent transformation, the design complexity of optimization problem is reduced effectively. Thus, it is possible to obtain the desired result even with a not so large number of training samples. Meanwhile, we yield 2000 testing samples to obtain the predicted results at the online prediction phase. The proposed ``ISACNN" has 3 CL layers with 2, 4, 8 filters and kernels size $(5,1)$, $(3,1)$, $(3,1)$, respectively.  To train the proposed ``ISACNN", the learning rate is initialized at 0.001 and the Adam optimizer is adopted with the maximum number of epoches being 500 and mini-batch size being 256. Besides, the EarlyStopping with patience 20 is used to prevent overfitting for enhancing the training efficiency, while ReduceLROnPlateau with patience 10 and factor 0.33 is applied to update the learning rate for accelerating the convergence.

First, we exhibit the training and testing performance with different NNs in Fig. \ref{Fig_train}, where ``Train\_num" in the legend denotes the number of training samples, the ``Train\_loss" and ``Val\_loss" in the y-label denote the training loss in the training sample set and the validation loss in the validation sample set, respectively. The baseline \emph{FC-NN} is a classical full-connected neural network \cite{C22}, which is mainly consists of four FC layers respectively with $8N_t, 4N_t, 2N_t$ and $N_t$ neurons. Like \emph{ISACNN}, the BN layer is also adopted before each FC layer in \emph{FC-NN} for better training. It is seen from Fig. \ref{Fig_train} that in the case of no overfitting, the more training samples, the less epoches required for convergence, and the better the training effect. Moreover, it is worthy noticing that even in the case of small sample training (such as train\_num $= 1000, 2000, 5000$), \emph{ISACNN} performs well on both the training set and the verification set, while the training loss curve and the validation loss curve for \emph{FC-NN} both oscillate violently. This indicates that the proposed \emph{ISACNN} is a high-powered NN with strong ability of data feature extraction, which is very suitable for ISAC systems.

\begin{figure}[h]
  \centering
  \includegraphics[width=0.45\textwidth]{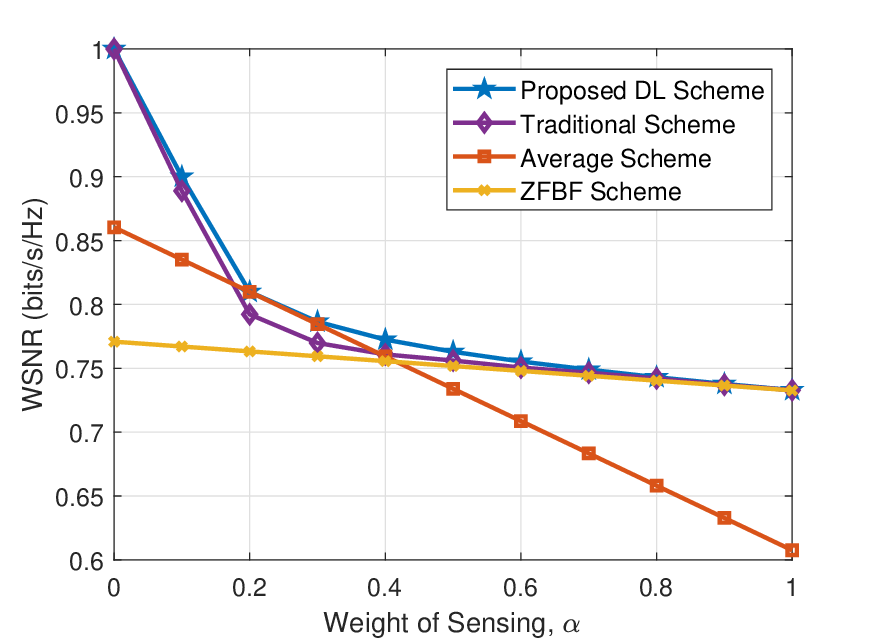}
  \caption{The performance comparison of different schemes under different weight of sensing.}
  \label{Fig_compare}
\end{figure}

Further, we present the performance of the proposed DL-based scheme over baseline schemes in Fig. \ref{Fig_compare}, i.e., \emph{Traditional Scheme} with AO method and the use of successive convex approximation (SCA), \emph{Average Scheme} with $\sigma_{s,i}=\frac{P_s}{N_t}, \forall i$, and \emph{ZFBF Scheme} with zero-forcing communication receive beamforming and SIC [14].  It can be seen that the weighted sum of normalized sensing rate and normalized communication rate (WSNR) of the proposed DL-based scheme achieves 1 on the point of $\alpha=0$ (namely only for communication), which means the solution obtained by the proposed DL scheme is equal to the maximum communication rate $M_c$.  In other words, this solution is optimal and our proposed DL-based scheme is effective. Moreover, the WSNR of three schemes all decreases as the weight of sensing $\alpha$ increases. This is because when the weight of sensing $\alpha$ increases, the system optimization focuses more on the sensing rate maximization, while the interference of CUs is inevitable for sensing performance, resulting in the decrease of the WSNR. This can also be verified from the point $\alpha=1$ (namely only for sensing) that the performance of the proposed DL-based scheme, the traditional scheme and the ZFBF scheme are all the same.  Besides, it is found that the performance of the Average scheme is worse than that of the ZFBF scheme at low sensing weight region but better than the ZFBF scheme at high sensing weight region, and the proposed DL-based scheme always performs best in the whole region of $\alpha$. Although the performance gap between the traditional scheme and the proposed DL-based scheme is small when $\alpha$ is within the range of $[0,0.1]$ and $[0.6,1]$, the traditional scheme requires two layers of iteration (using SCA technique to handle non-convexity and using AO method to handle coupled optimization variables), whose computational complexity is significantly higher than that of the proposed DL-based scheme. That also verifies the superiority of proposed DL-based scheme.

\begin{figure}[h]
  \centering
  \includegraphics[width=0.45\textwidth]{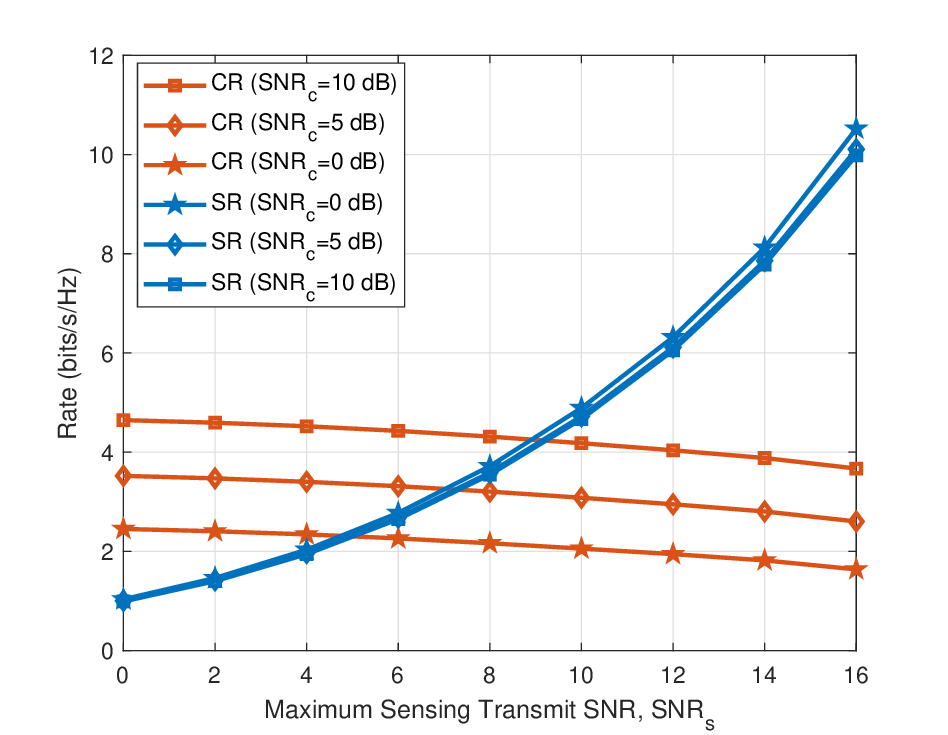}
  \caption{The performance versus the maximum sensing transmit SNR under different communication transmit SNR.}
  \label{Fig_power}
\end{figure}

\begin{figure}[h]
  \centering
  \includegraphics[width=0.45\textwidth]{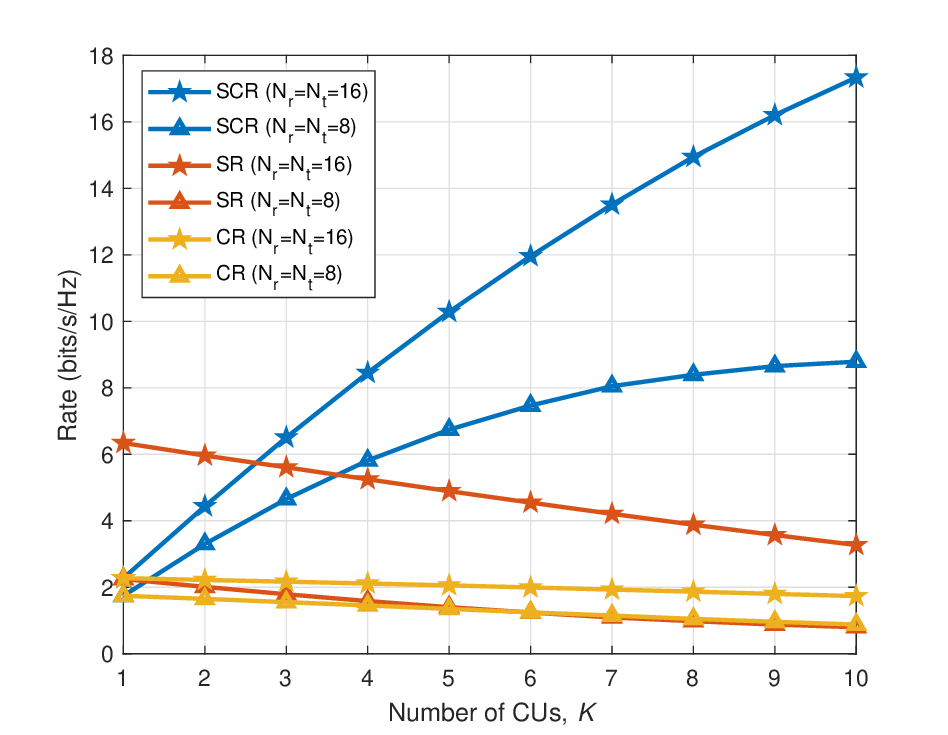}
  \caption{The performance versus the numbers of CUs with different BS antennas.}
  \label{Fig_K_N}
\end{figure}

\begin{figure*}
  \centering
  \includegraphics[width=1\textwidth]{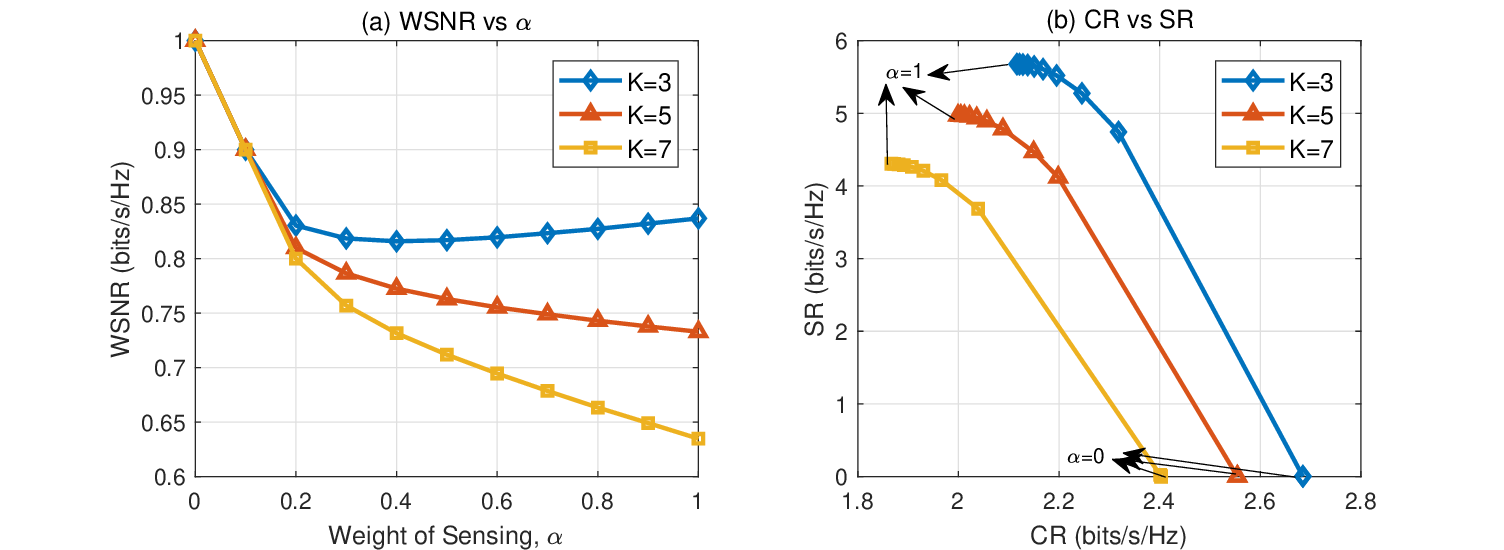}
  \caption{The performance versus the weight of sensing with different numbers of CUs.}
  \label{Fig_alpha_K}
\end{figure*}
Then, we investigate the impacts of the sensing SNR and the communication SNR on the performance of the proposed DL-based scheme. It is seen from Fig. \ref{Fig_power} that the sensing rate (SR) increases while the communication rate (CR) decreases as the maximum sensing transmit power increases. This is because more sensing transmit power brings in more MI on sensing, but leads to more interference on communication. Conversely, it is observed that the case with more communication transmit power leads to more communication rate and less sensing rate. Moreover, it is found that the increment of communication transmit power produces more influence on the communication performance but less effect on the sensing performance. Similarly, the change of sensing transmit power has a great impact on sensing performance but affect a little on communication performance. Therefore, the desired performance of ISAC systems can be achieved by adjusting the sensing transmit power and the communication transmit power.

In Fig. \ref{Fig_K_N}, we study the influence of the number of BS antennas and the number of CUs on the sensing rate (SR) and communication rate (CR) of the proposed DL-based scheme. It is observed that the sum of communication rate (SCR) increases while the sensing rate and the communication rate both decrease with the increase of number of CUs. This is because more CUs means more communication interference. On the one hand, the more communication-to-sensing interference leads to the decrease of sensing rate. On the other hand, more inter-user interference degrades the performance of communication. Moreover, as the number of BS antennas increases, the sensing rate and the communication rate both increase substantially thanks to the gain provided by more antennas. However, the gain obtained by increasing the number of BS antennas is limited and the addition of BS antennas causes relatively large overhead. Thus, it makes sense to resist interference by properly increasing the number of BS transmit/receive antennas in the ISAC systems.

Next, we show the impact of the weight of sensing and the number of CUs on the performance of the system performance. As is seen from Fig. \ref{Fig_alpha_K} (a), when the weight of sensing $\alpha$ is less than or equal to 0.4, the WCSR decreases with the increase of $\alpha$. When $\alpha$ is greater than 0.4, the WCSR in the case of fewer CUs ($K=3$) increases slowly, but decreases in the case of more CUs ($K= 5,7$), and the more CUs, the faster the decline. This is because the sensing rate maximization and the communication rate maximization are competitive rather than mutually beneficial relations under limited resources. When the number of CUs is small, it can be known that the lowest point of the WSNR curve is the maximum competition between sensing and communication. However, when there are a large number of CUs, the performance of sensing and communication both are seriously degraded, resulting in a monotonous decline of WSNR curve. From Fig. \ref{Fig_alpha_K} (b), we can know that the weight of sensing $\alpha$ has a great impact on the sensing rate and communication rate at the region of [0, 0.5], but a small impact at the region of [0.5, 1]. This is because when the optimization focuses on communication, the overall performance can be improved by adjusting the sensing transmit power, while when the optimization focuses on sensing, the impact of communication interference greatly limits the improvement of performance. Thus, choosing a suitable sensing weight is critical to the ISAC performance.

\begin{figure}[h]
  \centering
  \includegraphics[width=0.45\textwidth]{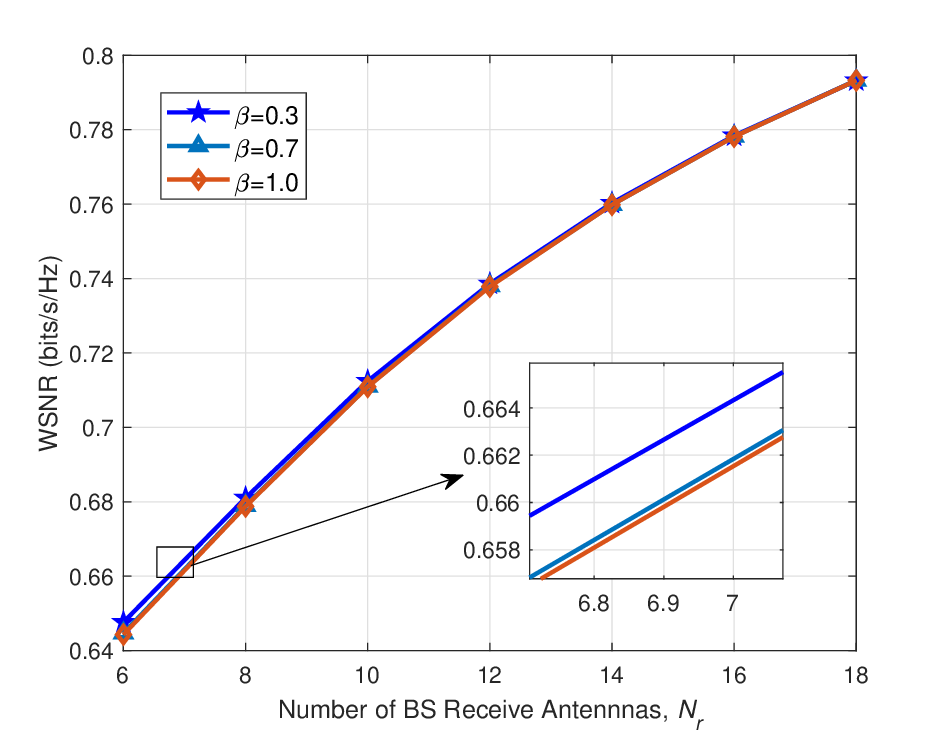}
  \caption{The performance versus the number of BS receive antenna under imperfect CSI.}
  \label{Fig_CSI}
\end{figure}
Finally, we consider the case of imperfect CSI, where the communication channel estimation is not perfect, which fits most practical systems. Here, we use $\beta \in [0,1]$ to denote the CSI accuracy. In order to characterize the communication CSI, we adopt the commonly used imperfect CSI model \cite{CSI}, i.e., $\mathbf{H}=\sqrt{\beta}\mathbf{\hat{H}}+\mathbf{E}$, where $\mathbf{H}$ is the real CSI, $\mathbf{\hat{H}}$ is the estimated CSI and $\mathbf{E}\sim \mathcal{CN}(0,(1-\beta)\mathbf{I})$ is the channel estimation error matrix. It is assumed that the estimated CSI and the real estimated are acquired by practical channel estimator and ideal channel estimator, respectively. During the offline training process, the estimated CSI is fed into ISACNN and the real CSI is used to calculate the training loss for unsupervised learning. On the online prediction process, only the estimated CSI is required to obtain the desired solution. Comparing the cases ($\beta=0.3, 0.7$) and the case with perfect CSI ($\beta=1$) in Fig. \ref{Fig_CSI}, it is seen that the performance loss caused by channel estimation error almost can be ignored. This is because the ISACNN learns the relationship between the real CSI and the estimated CSI in the training process, and presents its robustness in the prediction process. This also reflects the advantage of DL-based scheme for uncertainty optimization compared with the traditional optimization methods.  Moreover, it is found that the performance gap decreases as the number of BS receive antennas increases, which means the antennas gain can effectively compensate for the performance loss due to imperfect channel estimation in ISAC systems.

The above simulation results demonstrate that the proposed DL-based scheme has several advantages over traditional optimization schemes in terms of both system performance and computational complexity.  Moreover, it outperforms baselines consistently and is effective even with limited training samples.  The trade-offs between sensing and communication rates are observed when adjusting transmit powers and function weight.  Increasing the number of base station antennas improves performance, although there are diminishing returns.  The proposed DL-based scheme also shows robustness to imperfect CSI, and additional receive antennas can help compensate for this imperfection.  Overall, these findings validate the feasible, effectiveness and robustness of the proposed DL-based scheme for ISAC systems.

\section{Conclusion}
This paper provided a novel DL-based uplink ISAC design framework for 6G wireless networks. To mitigate the mutual interference and enhance the overall system performance, a joint sensing transmit waveform and communication receive beamforming design for the weighted sum of normalized sensing rate and normalized communication rate maximization was put forward. The design was formulated as a complicated non-convex optimization problem, which is not suitable to be directely solved by traditional optimization methods. To this end, we implemented a series of equivalent transformation of the original problem to reduce the complexity and provided a DL-based scheme to obtain a feasible solution. The effectiveness and robustness of the proposed DL-based scheme were verified by both theoretical analysis and numerical simulation.

\begin{appendices}
\section{The Proof of Lemma 1}\label{proof1}
For ease of derivation, we define $\mathbf{b}_i$ as the $i$-th column vector of $\mathbf{B}$, $\mathbf{c}_j$ as the $j$-th column vector of $\mathbf{C}$ and $c_{i,j}$ as the element in $i$-th low, $j$-th column of $\mathbf{C}$. Based on the definition of Kronecker product, we can obtain
\begin{eqnarray}
  ({{\mathbf{C}}^{T}}\otimes \mathbf{A})\text{vec}\left( \mathbf{B} \right)&=&\left[ \begin{matrix}
   {{c}_{11}}\mathbf{A} & {{c}_{21}}\mathbf{A} & \cdots  & {{c}_{p1}}\mathbf{A}  \\
   {{c}_{12}}\mathbf{A} & {{c}_{21}}\mathbf{A} & \cdots  & {{c}_{p2}}\mathbf{A}  \\
   \vdots  & \vdots  & \vdots  & \vdots   \\
   {{c}_{1q}}\mathbf{A} & {{c}_{21}}\mathbf{A} & \cdots  & {{c}_{pq}}\mathbf{A}  \\
\end{matrix} \right]\left[ \begin{matrix}
   {{\mathbf{b}}_{1}}  \\
   {{\mathbf{b}}_{2}}  \\
   \vdots   \\
   {{\mathbf{b}}_{p}}  \\
\end{matrix} \right] \nonumber \\
&=&\left[ \begin{matrix}
   \mathbf{A}\sum\limits_{j=1}^{p}{{{c}_{j1}}{{\mathbf{b}}_{1}}}  \\
   \mathbf{A}\sum\limits_{j=1}^{p}{{{c}_{j2}}{{\mathbf{b}}_{2}}}  \\
   \vdots   \\
   \mathbf{A}\sum\limits_{j=1}^{p}{{{c}_{j}}{{\mathbf{b}}_{p}}}  \\
\end{matrix} \right]=\left[ \begin{matrix}
   \mathbf{AB}{{\mathbf{c}}_{1}}  \\
   \mathbf{AB}{{\mathbf{c}}_{2}}  \\
   \ldots   \\
   \mathbf{AB}{{\mathbf{c}}_{q}}  \\
\end{matrix} \right] \nonumber\\
&=&\text{vec}(\mathbf{ABC}).
\end{eqnarray}
The proof is completed.

\section{The Proof of Lemma 2}\label{proof2}
According to the properties of the Kronecker product and the matrix determinant, we have
\begin{eqnarray}
&&\left| {{\mathbf{I}}_{mn}}+\mathbf{AB}\otimes \mathbf{CD} \right| \nonumber \\
&=&\left| {{\mathbf{I}}_{mn}}+\left( \mathbf{A}\otimes \mathbf{C} \right)\left( \mathbf{B}\otimes \mathbf{D} \right) \right|\nonumber\\
&=&1\times \left| \begin{matrix}
   {{\mathbf{I}}_{mn}}+\left( \mathbf{A}\otimes \mathbf{C} \right)\left( \mathbf{B}\otimes \mathbf{D} \right) & \mathbf{0}  \\
   \mathbf{B}\otimes \mathbf{D} & {{\mathbf{I}}_{mn}}  \\
\end{matrix} \right|\times 1 \nonumber\\
&=&\left| \begin{matrix}
   {{\mathbf{I}}_{mn}} & -\mathbf{A}\otimes \mathbf{C}  \\
   \mathbf{0} & {{\mathbf{I}}_{mn}}  \\
\end{matrix} \right|\left| \begin{matrix}
   {{\mathbf{I}}_{mn}}+\left( \mathbf{A}\otimes \mathbf{C} \right)\left( \mathbf{B}\otimes \mathbf{D} \right) & \mathbf{0}  \\
   \mathbf{B}\otimes \mathbf{D} & {{\mathbf{I}}_{mn}}  \\
\end{matrix} \right|\times 1\nonumber\\
&=&\left| \begin{matrix}
   {{\mathbf{I}}_{mn}} & -\mathbf{A}\otimes \mathbf{C}  \\
   \mathbf{B}\otimes \mathbf{D} & {{\mathbf{I}}_{mn}}  \\
\end{matrix} \right|\left| \begin{matrix}
   {{\mathbf{I}}_{mn}} & \mathbf{A}\otimes \mathbf{C}  \\
   \mathbf{0} & {{\mathbf{I}}_{mn}}  \\
\end{matrix} \right|\nonumber \\
&=&\left| \begin{matrix}
   {{\mathbf{I}}_{mn}} & \mathbf{0}  \\
   \mathbf{B}\otimes \mathbf{D} & \left( \mathbf{B}\otimes \mathbf{D} \right)\left( \mathbf{A}\otimes \mathbf{C} \right)+{{\mathbf{I}}_{mn}}  \\
\end{matrix} \right|\nonumber\\
&=&\left| {{\mathbf{I}}_{mn}}+\left( \mathbf{B}\otimes \mathbf{D} \right)\left( \mathbf{A}\otimes \mathbf{C} \right) \right|\nonumber\\
&=&\left| {{\mathbf{I}}_{mn}}+\mathbf{BA}\otimes \mathbf{DC} \right|.
\end{eqnarray}
The proof is completed.

\section{The Derivations of $M_s$ and $M_c$ }
We first derive the maximum sensing rate $M_s$. It is known that the maximum sensing rate can be achieved when there not exists any CU. In this context, $\sigma_{h,j}=\frac{1}{\sigma_n^{2}}, \forall j$. Thus, the optimization problem for sensing rate maximization can be formulated as
\begin{eqnarray}
\max_{\bm{\sigma}_s} &&\frac{N_r}{L}\sum_{i=1}^{N_t}\log_2 (1+\frac{\sigma_{t,i}}{\sigma_{n}^2}\sigma_{s,i})  \label{OP_Ms},  \\
\text { s.t. } && \sum_{i=1}^{N_t}{\sigma_{s, i}}\leq P_s. \nonumber
\end{eqnarray}
For this problem, we can make use of the well-known water-filling strategy to effectively obtain the solution of $\sigma_{s,i}$ \cite{Waterfilling}. Specifically, we construct the Lagrange function of $\sigma_{s,i}$ as
\begin{equation}
	\mathcal{L}_s(\sigma_{s,i})=\frac{N_r}{L}\sum_{i=1}^{N_t}\log_2 (1+\frac{\sigma_{t,i}}{\sigma_{n}^2}\sigma_{s,i})+\mu(P_s-\sum_{i=1}^{N_t}{\sigma_{s, i}}).
\end{equation}
Then, let the first derivative of $\mathcal{L}_s(\sigma_{s,i})$ be $0$, we have
\begin{equation}\label{KKT1} \sigma_{s,i}=\frac{L}{N_r\mu}-\frac{\sigma_n^{2}}{\sigma_{t,i}}=\hat{\mu}-\frac{\sigma_n^{2}}{\sigma_{t,i}}.
\end{equation}
 By taking (\ref{KKT1}) into the condition $P_s=\sum_{i=1}^{N_t}{\sigma_{s, i}}$,  we can obtain the solution for the maximum sensing rate as
\begin{equation} \sigma_{s,i}^{\text{opt}}=\max(\hat{\mu}-{\sigma_n^{2}}/{\sigma_{t,i}},0),
\end{equation}
where $\hat{\mu}=(P_s+\sum_{i=1}^{N_t} {\sigma_n^{2}}/{\sigma_{t,i}})/N_t$. As a result, the maximum sensing rate is given by
\begin{equation}\label{M_s}
  M_s=\frac{N_r}{L}\sum_{i=1}^{N_t}\log_2 (1+{\sigma_{t,i}}\sigma_{s,i}^{\text{opt}}/\sigma_{n}^2).
\end{equation}
Then, for the maximum communication rate $M_c$, it is intuitive that $M_c$ can be achieved when no sensing signal is transmitted, i.e., $\sigma_{s,i}=0, \forall i$. Take it into (\ref{newRc}), the maximum communication rate is given by
\begin{equation}\label{M_c}
  M_c=\frac{1}{K}\sum_{k=1}^{K}\log_{2}(1+{p_k\mathbf{h}_k^T\mathbf{\hat{R}}_k^{-1}\mathbf{h}_k^*}),
\end{equation}
where $\mathbf{\hat{R}}_k=\sum_{i \neq k}^{K} p_{i} \mathbf{h}_{i}^{*}\mathbf{h}_{i}^T+\sigma_n^{2}\mathbf{I}_{N_r}$.

\end{appendices}


\begin{thebibliography}{1}
\bibitem{C0}
Q. Qi, X. Chen, C. Zhong, C. Yuen, and Z. Zhang, ``DL-based joint waveform and beamforming design for integrated sensing and communication," in \emph{Proc. IEEE GC Wkshps}, Dec. 2023, pp. 1-6.

\bibitem{C1}
Z. Zhang, Y. Xiao, Z. Ma, M. Xiao, Z. Ding, X. Lei, G. K. Karagiannidis, and P. Fan, ``6G wireless networks: Vision, requirements, architecture, and key technologies," \emph{IEEE Veh. Technol. Mag.}, vol. 14, no. 3, pp. 28-41, Sep. 2019.


\bibitem{C2}
F. Liu,  Y. Cui , C. Masouros, J. Xu, T. X. Han, Y. C. Eldar, and S. Buzzi, ``Integrated sensing and communications: Towards dual-functional wireless networks for 6G and beyond", \emph{ IEEE J. Sel. Areas Commun.	},  vol. 40, no. 6, pp. 1728-1767, Jun. 2022.

\bibitem{C3}
D. K. P. Tan, J. He, Y. Li, A. Bayesteh, Y. Chen, P. Zhu, and W. Tong, ``Integrated sensing and communication in 6G: Motivation, use cases, requirements, challenges and future directions," in \emph{Proc. IEEE Inter. Symp. Joint Commun. Sensing (JC\&S)}, Feb. 2021, pp. 1-6.


\bibitem{C4}
Q. Zhang, H. Sun, X. Gao, X. Wang, and Z. Feng, ``Time-division ISAC enabled connected automated vehicles cooperation algorithm design and performance evaluation," \emph{IEEE J. Sel. Areas Commun.}, vol. 40, no. 7, pp. 2206-2218, Jul. 2022.

\bibitem{C5}
M. Temiz, E. Alsusa, and M. W. Baidas, ``A dual-function massive MIMO uplink OFDM communication and radar architecture," \emph{IEEE Trans. Cognit. Commun. Networking}, vol. 8, no. 2, pp. 750-762, Jun. 2022.

\bibitem{C6}
X. Chen, Z. Feng, Z. Wei, P. Zhang, and X. Yuan, ``Code-division OFDM joint communication and sensing system for 6G machine-type communication," \emph{IEEE Internet Things J.}, vol. 8, no. 15, pp. 12093-12105, Aug. 2021.

\bibitem{C7}
X. Mu, Z. Wang, and Y. Liu,  ``NOMA for integrating sensing and communications towards 6G: A multiple access perspective," [Online]: https://arxiv.org/abs/2206.00377, 2022.

\bibitem{C8}
F. Liu, L. Zhou, C. Masouros, A. Li, W. Luo, and A. Petropulu, ``Toward dual-functional radar-communication systems: Optimal waveform
design," \emph{IEEE Trans. Signal Process.}, vol. 66, no. 16, pp. 4264-4279, Jun. 2018.


\bibitem{C9}
Z. Wang, Y. Liu, X. Mu, Z. Ding, and O. A. Dobre, ``NOMA empowered integrated sensing and communication," \emph{IEEE Commun. Lett.}, vol. 26, no. 3, pp. 677-681, Mar. 2022.

\bibitem{C10}
X. Mu, Y. Liu, L. Guo, J. Lin, and L. Hanzo, ``NOMA-aided joint radar and multicast-unicast communication systems," \emph{IEEE J. Sel. Areas Commun.}, vol. 40, no. 6, pp. 1978-1992, Jun. 2022.

\bibitem{C11}
Z. Zhang, H. Sun, and R. Q. Hu, ``Downlink and uplink non-orthogonal multiple access in a dense wireless network," \emph{IEEE J. Sel. Areas Commun.}, vol. 35, no. 12, pp. 2771-2784, Dec. 2017.

\bibitem{C12}
X. Chen, Z. Zhang, C. Zhong, and D. W. K. Ng, ``Exploiting multiple-antenna techniques for non-orthogonal multiple access," \emph{IEEE J. Sel. Areas Commun.}, vol. 35, no. 10, pp. 2207-2220, Oct. 2017.


\bibitem{C13}
C. Ouyang, Y. Liu, and H. Yang, ``On the performance of uplink ISAC systems," \emph{IEEE Commun. Lett.}, vol. 26, no. 8, pp. 1769-1773, Aug. 2022.

\bibitem{C14}
Z. Wang, X. Mu, Y. Liu, X. Xu, and P. Zhang, ``NOMA-aided joint communication, sensing, and multi-tier computing systems," [Online]: https://arxiv.org/abs/2205.08272, 2022.

\bibitem{C15}
A. Tajer, G. Jajamovich, and X. Wang, ``Optimal joint target detection and parameter estimation by MIMO radar," \emph{IEEE J. Sel. Topics Signal Process.}, vol. 4, no. 1, pp. 127-145, Feb. 2010.

\bibitem{C16}
Y. Fu and Z. Tian, ``Cramer-Rao bounds for hybrid TOA/DOA-based location estimation in sensor networks," \emph{IEEE Signal Process. Lett.}, vol. 16, no. 8, pp. 655-658, Aug. 2009.

\bibitem{C17}
H. L. V. Trees and K. L. Bell, \emph{Bayesian bounds for parameter estimation and nonlinear filtering/tracking}. New York: Wiley-Intersci.,
2007.

\bibitem{C18}
B. Tang and J. Li, ``Spectrally constrained MIMO radar waveform design based on mutual information," \emph{IEEE Trans. Signal Process.}, vol. 67, no. 3, pp. 821-834, Feb. 2019.

\bibitem{C19}
S. Shi, Z. He, Q. He, and Z. Cheng, ``Co-design for MU-MIMO communication and MIMO radar systems based on mutual information," in \emph{Proc. 2022 IEEE Radar Conference}, 2022, pp. 1-6.

\bibitem{C20}
X. Wang, Z. Fei, J. Huang, and H. Yu, ``Joint waveform and discrete phase shift design for RIS-assisted integrated sensing and communication system under Cramer-Rao bound constraint," \emph{IEEE Trans. Veh. Technol.}, vol. 71, no. 1, pp. 1004-1009, Jan. 2022.

\bibitem{DL1}
H. Lee, S. H. Lee, T. Q. S. Quek, and I. Lee, ``Deep learning framework for wireless systems: Applications to optical wireless communications," \emph{IEEE Commun. Mag.}, vol. 57, no. 3, pp. 35-41, Mar. 2019.

\bibitem{DL2}
G. Zhu, D. Liu, Y. Du, C. You, J. Zhang, and K. Huang, ``Towards an intelligent edge: Wireless communication meets machine learning," \emph{IEEE Commun. Mag.}, vol. 58, no. 1, pp. 19-25, Jan. 2020.


\bibitem{C21}
Y. Zhang, J. Sun, J. Xue, G. Y. Li, and Z. Xu, ``Deep expectation-maximization for joint MIMO channel estimation and signal detection," \emph{IEEE Trans. Signal Process.}, vol. 70, pp. 4483-4497, 2022.

\bibitem{C22}
T. Lin and Y. Zhu, ``Beamforming design for large-scale antenna arrays using deep learning," \emph{IEEE Wireless Commun. Lett.}, vol. 9, no. 1, pp. 103-107, Jan. 2020.

\bibitem{C23}
J. Tian, Q. Liu, H. Zhang, and D. Wu, ``Multiagent deep-reinforcement-learning-based resource allocation for heterogeneous QoS guarantees for vehicular networks,"\emph{IEEE Internet  Things J.}, vol. 9, no. 3, pp. 1683-1695, Feb. 2022.

\bibitem{Waterfilling}
G. Scutari, D. P. Palomar, and S. Barbarossa, ``The MIMO iterative waterfilling algorithm," \emph{IEEE Trans. Signal Process.}, vol. 57, no. 5, pp. 1917-1935, May 2009.

\bibitem{complexity1}
P. Molchanov, S. Tyree, T. Karras, T. Aila, and J. Kautz, ``Pruning convolutional neural networks for resource effcient inference," [Online]. Available: https://arxiv.org/abs/1611.06440.

\bibitem{complexity2}
W. Xia, G. Zheng, Y. Zhu, J. Zhang, J. Wang, and A. P. Petropulu, ``A deep learning framework for optimization of MISO downlink beamforming," \emph{IEEE Trans. Commun.}, vol. 68, no. 3, pp. 1866-1880, Mar. 2020.

\bibitem{pathlossmodel}
3GPP, ``Coordinated multi-point operation for LTE physical layer aspects (Rel. 11)," Feb. 2011.

\bibitem{RT}
B. Tang, M. M. Naghsh, and J. Tang, ``Relative entropy-based waveform design for MIMO radar detection in the presence of clutter and interference," \emph{IEEE Trans. Signal Process.}, vol. 63, no. 14, pp. 3783-3796, Jul. 2015.

\bibitem{CSI}
Q. Qi, X. Chen, L. Lei, C. Zhong and Z. Zhang, ``Outage-constrained robust design for sustainable B5G cellular internet of things," \emph{IEEE Trans. Wireless Commun.}, vol. 18, no. 12, pp. 5780-5790, Dec. 2019.

\end{thebibliography}
\end{document}